\documentclass[%
 reprint,
 onecolumn,
 amsmath,amssymb,
 aps,
]{revtex4-2}

\usepackage{graphicx}
\usepackage{dcolumn}
\usepackage{bm}

\begin{document}

\preprint{APS/123-QED}

\title{Asymmetric rectified electric and concentration fields in multicomponent electrolytes with surface reactions$^\dag$}

\author{Nathan Jarvey\textit{$^{a}$}}
\author{Filipe Henrique\textit{$^{a}$}}%
\author{Ankur Gupta \textit{$^{*a}$}}

\affiliation{a: Department of Chemical and Biological Engineering, University of Colorado, Boulder}

\date{\today}

\begin{abstract}
Recent experimental studies have utilized AC electric fields and electrochemical reactions in multicomponent electrolyte solutions to control colloidal assembly. However, theoretical investigations have thus far been limited to binary electrolytes and have overlooked the impact of electrochemical reactions. In this study, we address these limitations by analyzing a system with multicomponent electrolytes, while also relaxing the assumption of ideally blocking electrodes to capture the effect of surface electrochemical reactions. Through a regular perturbation analysis in the low-applied-potential regime, we solve the Poisson-Nernst-Planck equations and obtain effective equations for electrical potential and ion concentrations. By employing a combination of numerical and analytical calculations, our analysis reveals a significant finding: electrochemical reactions alone can generate asymmetric rectified electric fields (AREFs), i.e., time-averaged, long-range electric fields, even when the diffusivities of the ionic species are equal. This finding expands our understanding beyond the conventional notion that AREFs arise solely from diffusivity contrast. Furthermore, we demonstrate that AREFs induced by electrochemical reactions can be stronger than those resulting from asymmetric diffusivities. Additionally, we report the emergence of asymmetric rectified concentration fields (ARCFs), i.e., time-averaged long-range concentration fields, which supports the electrodiffusiophoresis mechanism of colloidal assembly observed in experiments. We also derive analytical expressions for AREFs and ARCFs, emphasizing the role of imbalances in ionic strength and charge density, respectively, as the driving forces behind their formation. The results presented in this article advance the field of colloidal assembly and also have implications for improved understanding of electrolyte transport in electrochemical devices.
\end{abstract}

\maketitle

\section{Introduction}
\label{sec: intro}
\par{}  Colloidal particles immersed in an electrolyte aggregate along a plane near an electrode when an AC field is applied
\cite{sides2001electrohydrodynamic, ristenpart2007electrohydrodynamic, wirth2011single, wirth2013electrolyte, hoggard2008electrolyte, prieve_2-d_2010, woehl_electrolyte-dependent_2014, saini2016influence} due to attractive electrohydrodynamic flows between the particles \cite{ristenpart2004assembly, ristenpart2007electrohydrodynamic}. Intriguingly, Woehl et al. \cite{woehl2015bifurcation} and Bukosky and Ristenpart \cite{bukosky2015simultaneous} reported that the planar height at which colloids aggregate exhibits a bifurcation that depends on the electrolyte type and the frequency of the AC field. This bifurcation was particularly surprising because the particles levitate several diameters away from the electrode \cite{woehl2015bifurcation}.

\par{} Since this discovery, Hashemi et al. \cite{hashemi_oscillating_2018} demonstrated through direct numerical simulations of the Poisson-Nernst-Planck (PNP) equations for a binary electrolyte that diffusivity contrast between anions and cations induces a long-range steady electric field, also referred to as an asymmetric rectified electric field (AREF). The strength of the AREF is dependent on the diffusivity contrast and frequency \cite{hashemi_asymmetric_2019}, which consequently determines the electrophoretic force on the particle and the bifurcation height. This mechanism was experimentally validated in a study by Bukosky et al. \cite{bukosky_extreme_2019}. Hashemi et al. \cite{hashemi_asymmetric_2020} have also argued that AREFs can directly impact or even dominate flows from mechanisms such as induced-charged electrophoresis. 

\par{} While AREFs are able to recapitulate experimental observations, their direct numerical simulation requires high-order adaptive meshing \cite{hashemi_oscillating_2018, mirzadeh2014conservative}, which poses its own challenge. To this end, for a binary electrolyte, Hashemi et al.  \cite{hashemi2020perturbation}  performed a regular perturbation expansion on the PNP equations in the low-applied-potential limit and showed that AREFs appear at the second order in applied potential. Balu and Khair \cite{balu_thin_2021}, in contrast, performed a singular perturbation expansion in the thin-double-layer limit and demonstrated that AREFs are recovered at the second order in the  ratio between double layer and cell length. While both of these studies have furthered our understanding of AREFs, they are limited to binary electrolytes. Moreover, the aforementioned analyses on AREFs rely on the ideally blocking electrode approximation, which can be a limiting factor \cite{silvera_batista_controlled_2017, wang2022long, wang2022visualization}. Wang et al. \cite{wang2022long} note that \textit{``the AREF theory assumes no flux for all ions at the electrodes; essentially, it does not account for Faradaic reactions (electrochemistry), which will take place at frequencies below 1 kHz in water".} In their work, in addition to the AC electric field, Wang et al. \cite{wang2022long} employed water splitting reactions and experimentally demonstrated that colloidal aggregation occurs at the location where the pH of the solution is at its maximum. A similar influence of pH on colloidal aggregation was also reported by Rath et al. \cite{rath2021ph}, where the authors employed the electroreduction of para-benzoquinone while also including a variable steady component of the applied potential.
\par{} Given the increasing interest in utilizing electrochemical reactions for manipulating colloidal assembly, we generalize the regular perturbation analysis of Hashemi et al. \cite{hashemi_asymmetric_2020} in the small potential limit to multicomponent electrolytes while also relaxing the ideally blocking electrode assumption. We find that AREFs can also be induced solely through electrochemical reactions, even for symmetric diffusivities, and can be stronger than AREFs created by diffusivity contrast alone. This demonstrates that AREFs may be present in a wider parameter space than previously anticipated. In addition to AREFs, we report the formation of asymmetric rectified concentration fields (ARCFs). While AREFs induce an electrophoretic force, ARCFs induce a diffusiophoretic (or osmotic, depending on the definition) force \cite{prieve1984motion, anderson1989colloid, shin2016size, velegol2016origins, banerjee2019long, gupta2019diffusiophoretic, gupta2020diffusiophoresis, chu2022tuning, ganguly2023going, raj2023two}. We discover that ARCFs are primarily observed in systems with diffusivity contrast and that electrochemical reactions do not produce ARCFs, but can enhance  ARCFs caused by diffusivity asymmetry.
\par{} The simultaneous inclusion of AREFs and ARCFs could rationalize recent experimental findings. For instance, the colloidal aggregation reported in Wang et al. \cite{wang2022long} closely resembles the diffusiophoretic focusing for an acid-base reaction reported in Shi et al. \cite{shi2016diffusiophoretic} and Banerjee et al. \cite{banerjee2019long}, but as their system also includes an imposed electric field, the authors invoked the phenomena of electrodiffusiophoresis. Our  findings also provide mechanistic insights into the formation of AREFs and ARCFs. Specifically, we highlight that the imbalances in ionic strength and charge density lead to AREFs and ARCFs, respectively. We also provide convenient analytical expressions for AREFs and ARCFs, which although valid only in the limiting case of small applied potentials and thin electrical double layers, provide a good starting point for estimating their spatial variations.
\par{} We provide a simplified model for ease of understanding in Section \ref{sec: toy}. We outline the problem formulation in section \ref{sec: setup}. We perform a regular perturbation in the low-applied-potential limit and derive effective equations for AREFs and ARCFs; see sections \ref{sec: numerics} and \ref{sec: analytics}. Next, in section \ref{sec: results}, we validate our numerical results with analytical calculations and report the dependencies of AREFs on various parameters. We briefly discuss the factors which control the strength of ARCFs. Finally, we describe the limitations of our analysis, outline potential future research directions, and discuss the implications of our findings on colloidal assembly and electrochemical devices. 

\section{Toy Model}
\label{sec: toy}
\begin{figure}[ht!]
    \centering
    \includegraphics[width=0.45 \textwidth]{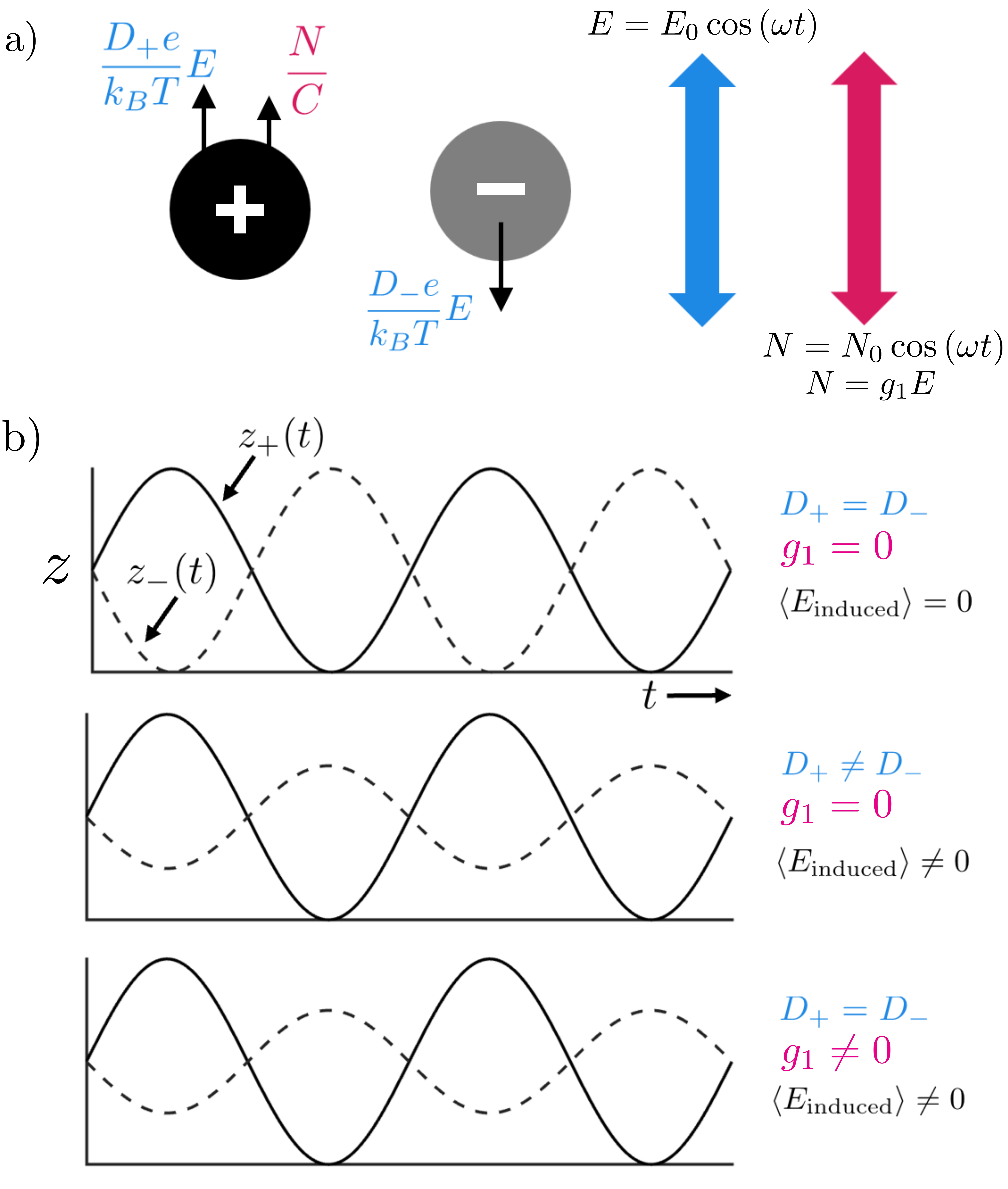}
    \caption{\textbf{Schematic of the toy model}. (a) A cation and an anion move in the $z$-direction as a response to an electric field $E$. The cation also moves in response to a surface redox flux $N$. Both $E$ and $N$ are sinusoidal in time. $N$ is proportional to $E$. The velocity induced by the electric field is dependent on the charge and the diffusivity of the ion. The velocity induced by surface reactions on the cation is proportional to the strength of the surface reaction. (b) The position of the cation ($z_+$) and anion ($z_-$) as a function of time $t$. When the diffusivities of each ion are equal and there are no reactions, the movement of ions has equal amplitude and thus no AREF is formed. If the diffusivities are unequal or surface reactions are present, the amplitudes are no longer equal and an AREF is formed.}
    \label{fig: toy}
\end{figure}
 Before delving into the details of the electrokinetic equations, inspired by Hashemi et al. \cite{hashemi_oscillating_2018, hashemi_asymmetric_2020}, we propose a toy model that is able to capture the effect of surface reactions. At the outset, we would like to clarify that the toy model described here ignores several complexities that are present in a real system. However, it provides a convenient choice to grasp the dominant physics within the system. 
\par{} We consider a system with a cation and an anion. Both the ions can move in the $z$-direction due to an applied sinusoidal electric field $E=E_0 \cos (\omega t)$, where $E_0$ is the amplitude, $\omega$ is the frequency and $t$ is time. In addition to the electric field, the cation also moves in the $z$-direction due to the presence of a redox flux field, denoted here as $N=N_0 \cos (\omega t) = g_1 E_0 \cos (\omega t)$; where $g_1$ is a constant. This assumption assumes that the redox flux is proportional to the electric field, which is valid for small amplitude oscillations \cite{prieve_2-d_2010}. Note that the redox flux is not consuming/producing the cation but is rather inducing a velocity on the cation; see Fig. \ref{fig: toy}. The two ions have valences of $+1$ and $-1$ and their locations are denoted $z_+$ and $z_-$, respectively. It is assumed that both the ions are at the location $z=0$ at $t=0$. The cation and anion have diffusivities of $D_+$ and $D_-$, respectively. 
\par{} The applied electric field is known to create an electromigrative flux, and the induced velocity for the cation and anion are given by $\pm \frac{D_{\pm} e}{k_B T} E_0 \cos (\omega t)$, where $e$ is the charge on an electron, $k_B$ is Boltzmann's constant, and $T$ is the temperature. The velocity induced on the cation due to the reactive flux is $\frac{g_1}{C} E_0 \cos (\omega t)$ (obtained by equating the convective flux to the redox flux), where $C$ is the concentration scale of the cations. This implies that one can write  
\begin{subequations}
    \begin{eqnarray}
    \frac{d z_+}{d t} =  \left( \frac{D_+ e}{k_B T} + \frac{g_1}{C} \right) E_0 \cos (\omega t), \\
     \frac{d z_-}{d t} =   - \frac{D_- e}{k_B T} E_0 \cos (\omega t),
    \end{eqnarray}
which upon integration yield $z_{\pm} = \pm m_{\pm} \frac{E_0}{\omega} \sin(\omega t)$, where $m_+ =  \left( \frac{D_+ e}{k_B T} + \frac{g_1}{C} \right)$ and $m_- = \frac{D_- e}{k_B T}$. The net electric field induced $E_{\textrm{induced}}$ by the ions at a location $z$ can be calculated by applying Coulomb's law, assuming the ions are point charges similar to Hashemi et al.\cite{hashemi_oscillating_2018}. For $|z| \gg z_{\pm}$ and time averaging, it is straightforward to obtain 
\begin{equation}
    \left< E_{\textrm{induced}} \right> \propto  E_0^2 \left( m_+^2 - m_-^2 \right),  
    \label{Eq: toy}
\end{equation}
where $\left< \right>$ represents time averaging and where we have ignored the higher order terms beyond the second order in $\left( \frac{z_{\pm}}{z} \right)^2$.
\par{} Clearly, as per Eq. (\ref{Eq: toy}), if $g_1=0$ and $D_+=D_-$, the induced electric field vanishes. If $g_1 = 0$ and $D_+ \neq D_-$, the induced electric field is an AREF and falls under the scenario described by Hashemi et al. \cite{hashemi_oscillating_2018, hashemi_asymmetric_2020}. However, even if $D_+ = D_-$, an AREF is also possible when $g_1 \neq 0$; see Fig. \ref{fig: toy}. This is the key finding that is explored in this paper, as a surface redox flux can also produce AREFs without the requirement of asymmetric diffusivities by enhancing the effective mobility of one ion. We reiterate that the toy model has its limitations, as it is not able to capture the subtle features that we discuss in the remainder of this manuscript.  
\end{subequations}

\section{Problem Setup}
\label{sec: setup}
\subsection{Dimensional Problem}
\label{sec: dimensional} 
\begin{figure*}[t!]
    \centering
    \includegraphics[width = 0.75 \textwidth]{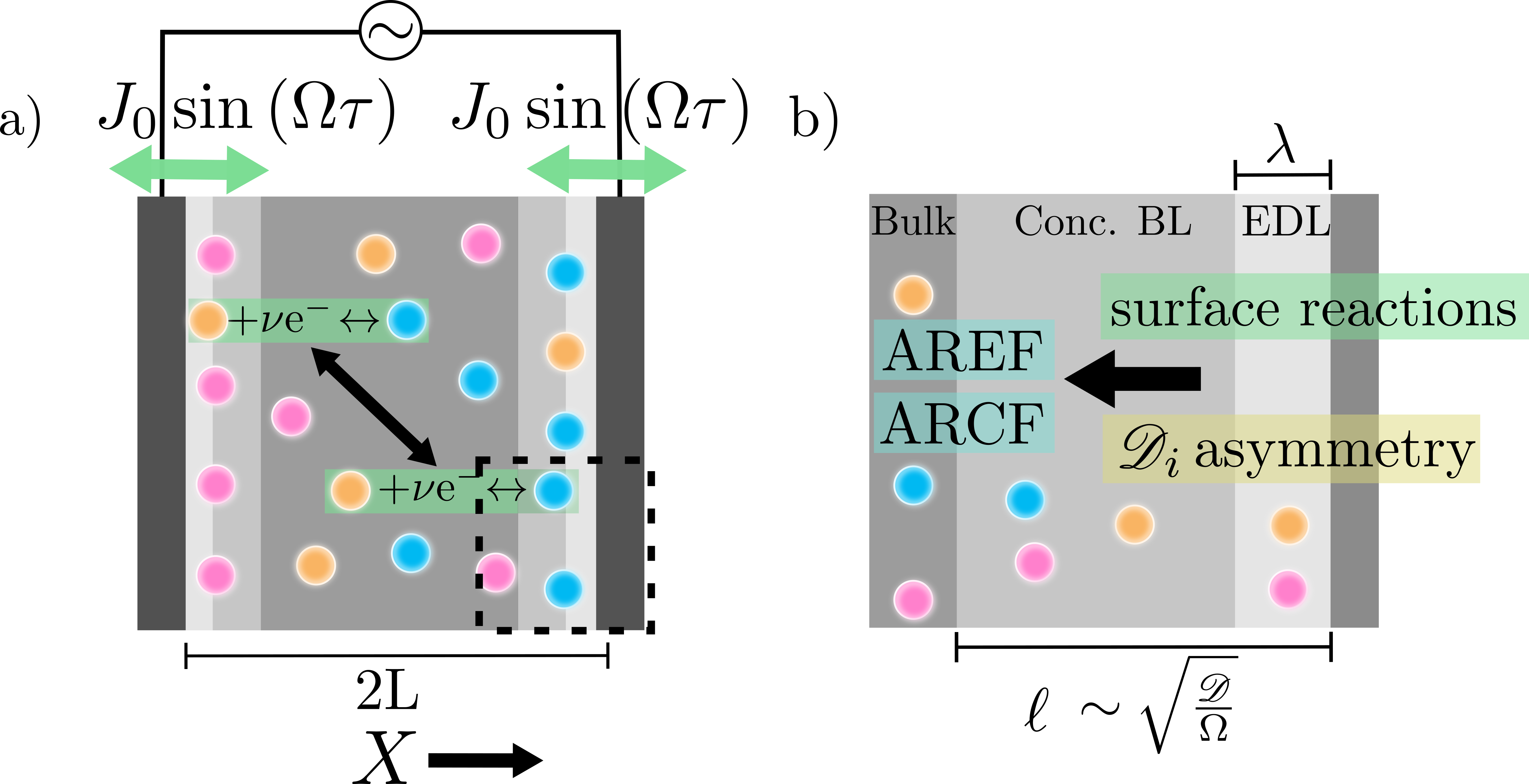}
    \caption{\textbf{Schematic of the model problem.} a) We consider a cell of length $2L$ with an arbitrary number of ions. $X$ is the spatial coordinate and $J_0 \sin{\left(\Omega \tau \right)}$ is the charge flux due to surface reactions. b) Zoomed-in schematic of the dashed box. The cell consists of three spatial regions: the electrical double layer (EDL), a concentration boundary layer (conc. BL), and the bulk. The thickness of the EDL is denoted by $\lambda$ and the thickness of the conc. BL is denoted by $\ell$. $\mathcal{D}_i$ is the diffusivity of the $i^{\textrm{th}}$ ion, $\mathcal{D}$ is a} characteristic diffusivity corresponding to the conc. BL length, and $\Omega$ is the frequency of the applied field. We show that $\mathcal{D}_i$ asymmetry and surface reactions can cause AREFs, and both 
    attributes do so due to an imbalance in ionic strength. We also show that $\mathcal{D}_i$ asymmetry can create an ARCF due to an imbalance in charge, and surface reactions can further enhance them.
    \label{Fig: Fig1}
\end{figure*}

We study a one-dimensional electrochemical system with an arbitrary number of ions and two electrodes separated by a distance $2L$; see Fig. \ref{Fig: Fig1}. An AC field of frequency $\Omega$ is applied to the electrochemical cell. $X = 0$ is at the centerline of the cell, $\ell$ is the length of the concentration boundary layer (conc. BL) and $\lambda$ is a measure of the length of the electrical double layer (EDL).  Here, we investigate the formation of AREFs and ARCFs in the presence of surface reactions without imposing any restrictions on ionic diffusivities. \par{} We seek to describe the spatial and temporal variations of ionic concentrations and potential in the system to subsequently determine the AREF and ARCF. Therefore, we write Poisson's equation \cite{deen2012analysis, bard2000electrochemical, newman2012electrochemical} 
\begin{subequations}
\label{Eq: governing_dim}
\begin{equation}
    \label{Eq: poisson}
    - \varepsilon \frac{\partial^2 \Phi}{\partial X^2} = Q_e,
\end{equation}
with $\varepsilon$ being the electrical permittivity of the solvent, $\Phi$ being the potential, $X$ being the spatial coordinate, and $Q_e$ being the volumetric charge density. $Q_e= \sum_i{e z_i C_i}$, where $e$ is the charge on an electron and $z_i$ and $C_i$ are the valence and concentration  of the $i^{\textrm{th}}$ ion, respectively. 
\par{} Ion transport is modeled using the Nernst-Planck equations \cite{deen2012analysis, bard2000electrochemical, newman2012electrochemical}, i.e., 
\begin{equation}
\frac{\partial C_i}{\partial \tau}  + \frac{\partial N_i}{\partial X} = 0,  
\label{Eq: nernst-planck}
\end{equation}
with $N_i$ as the flux of ion $i$, given by
\begin{equation}
N_i = - \mathcal{D}_i \frac{\partial C_i}{ \partial X} - \frac{\mathcal{D}_i z_i e C_i}{k_B T} \frac {\partial \Phi}{\partial X},     
\end{equation}
\end{subequations}
\noindent where $\tau$ is time, $\mathcal{D}_i$ is the diffusivity of the $i^{\textrm{th}}$ ion, $k_B$ is Boltzmann's constant, and $T$ is temperature. We note that Eq. (\ref{Eq: nernst-planck}) ignores any volumetric reactions. The charge flux (or current per unit area) is evaluated as $J = \sum_i z_i e N_i$.
\par{} Eqs. (\ref{Eq: governing_dim}) are subjected to the following boundary and initial conditions. The sinusoidal potential boundary conditions are given as 
\begin{subequations}
\label{Eq: bcic_dim}
\begin{equation}
    \Phi(\pm L, \tau) = \pm \Phi_D \sin{\left(\Omega \tau\right)}.
\end{equation}
The above equation ignores any native zeta potential of the electrodes, similar to Hashemi et al. \cite{hashemi2020perturbation}. We consider a surface reactive flux condition (i.e., non-ideally blocking) at the two electrodes 
\begin{equation}
N_i(\pm L,\tau)=  N_{i0} \sin{\left(\Omega \tau\right)}, 
\end{equation}
We note that the flux amplitude $N_{i0}$ may not be identical at the two electrodes \cite{jarvey2022ion}, but is assumed to be the same and time-independent for simplicity. Typically, $N_{i0}$ is dependent on applied potential and ion concentrations. We would like to clarify that that the applied flux is dependent on frequency and has the same sinusoidal dependency as the potential, i.e., it is assumed that the potential and fluxes are in-phase. The dependency of the amplitude $N_{i0}$ on $\Phi_D$ is discussed in Sections \ref{sec: asymptotic} and \ref{sec: limitation}. We also define current amplitude $J_0 = \sum_i z_i e N_{i0}$; see Fig. \ref{Fig: Fig1}. 
\par{} At $\tau = 0$, the electrical potential is
\begin{equation}
    \Phi(X,0) = 0, 
\end{equation}
and the concentrations are given by
\begin{equation}
    C_i (X,0) = C_{i0}.
\end{equation}
The initial concentrations are required to maintain electroneutrality, i.e., $\sum_i z_i C_{i0}=0$. 
\end{subequations}
\subsection{Dimensionless Problem}
\label{sec: nondimensional}
We write dimensionless concentration $c_i = \frac{C_i}{C^*}$, diffusivity $D_i = \frac{\mathcal{D}_i}{\mathcal{D}^*}$, time $t = \frac{\tau\mathcal{D}^*}{L^2}$, potential $\phi = \frac{e \Phi}{k_B T}$, spatial coordinate $x = \frac{X}{L}$, charge density $\rho_e = \frac{Q_e}{e C^*}$, charge flux $j = \frac{J L}{e \mathcal{D}^* C^*}$, species fluxes $n_i = \frac{N_i L}{\mathcal{D}^* C^*}$, and frequency $\omega = \frac{\Omega L^2}{\mathcal{D}^*}$, where $C^*$ is a reference concentration and $\mathcal{D}^*$ is a reference diffusivity. We also define  $\lambda=\sqrt{ \frac{\varepsilon k_B T}{e^2 C^*}}$ as a representative measure of double-layer thickness and $\kappa = \frac{L}{\lambda}$. We would like to clarify that $\lambda$ is not the true Debye length, as it is based on a reference value $C^{*}$ and not the ionic strength. We make this choice on purpose to decouple the effects of ionic valences and $\kappa$. As we will discuss later, the true Debye length is given by a combination of $\lambda$ and ionic strength. 
 
\par{} In dimensionless variables, Eqs. (\ref{Eq: governing_dim}) take the form
\begin{subequations}
\label{eq: nondimPNP}
\begin{equation}
    - \frac{\partial^2 \phi}{\partial x^2} = \kappa^2 \rho_e.
    \label{eq: nondimpoisson}
\end{equation}
\begin{equation}
    \frac{\partial c_i}{\partial t} - D_i \frac{\partial^2 c_i}{\partial x^2} - D_i z_i \frac{\partial}{\partial x} \left( c_i \frac{\partial \phi}{\partial x} \right) = 0,
    \label{eq: nondimNP}
\end{equation}
\end{subequations}
and Eqs. (\ref{Eq: bcic_dim}) become
\begin{subequations}
\label{eq: nondimBCs}
\begin{eqnarray}
\phi(\pm 1,t) = \pm \phi_D \sin{\left( \omega t\right)},\\
n_i(\pm 1, t) = n_{i0} \sin{\left( \omega t\right)}, \label{eq: nondimrxnBC}\\
\phi(x,0) = 0,\\
c_i(x,0) = c_{i0},
\end{eqnarray}
\end{subequations}
where $\phi_D = \frac{e \Phi_D}{k_B T}$, $n_i = - D_i \frac{\partial c_i}{\partial x} - D_i z_i c_i \frac{\partial \phi}{\partial x} $, $n_{i0} = \frac{N_{i0} L}{\mathcal{D}^* C^*}$, $j_0 = \frac{J_0 L}{e \mathcal{D}^* C^*}$ and $c_{i0} = \frac{C_{i0}}{C^*}$, while we require $\sum_{i}{z_i c_{i0}} = 0$ to maintain electroneutrality. 

\section{Asymptotic Solution for Small Applied Potentials}
\label{sec: asymptotic}
In this work, we employ a regular perturbation expansion in the low-applied-potential limit, i.e. $ \phi_D \ll 1$. While this limit is not directly observed in experiments (which generally tend to operate in moderate to large potential limits), it is able to capture the essential physics of the electrokinetic problem, albeit qualitatively \cite{woehl_electrolyte-dependent_2014, bukosky2015simultaneous, bukosky_extreme_2019}. This limit is also a common choice for theoretical developments \cite{hashemi2020perturbation, henrique2022charging, henrique2022impact, de1963porous, keh2000diffusiophoretic, hollingsworth2003broad, krozel1992electrostatic}. 
\par{} The perturbation expansions in powers of $\phi_D$ are $\phi = \phi^{(0)} + \phi_D \phi^{(1)} + \phi_D^2 \phi^{(2)} + O(\phi_D^3)$, $c_i = c_{i}^{(0)} + \phi_D c_{i}^{(1)} + \phi_D^2 c_i^{(2)} + O(\phi_D^3)$, $n_i = n_{i}^{(0)} + \phi_D n_{i}^{(1)} + \phi_D^2 n_i^{(2)} + O(\phi_D^3)$, $j = j^{(0)} + \phi_D j^{(1)} + \phi_D^2 j^{(2)} + O(\phi_D^3)$, and $\rho_e = \rho_e^{(0)} + \phi_D \rho_e^{(1)} + \phi_D^2 \rho_e^{(2)} + O(\phi_D^3)$,
where the superscripts (0), (1), and (2) refer to the leading-, first-, and second-order terms, respectively. 
\par{} In the small potential limit, we assume that the equilibrium cell potential is 0 and invoke the linearized Butler-Volmer kinetic equation to write $n_{i0} = \phi_D n_{i0}^{(1)}$. Thus, neglecting the equilibrium potential implies that the applied potential is the overpotential and the fluxes (and consequently the current) are proportional to the overpotential. This relationship was systematically derived by Prieve et al. \cite{prieve_2-d_2010}.  We acknowledge that this assumption ignores the impact of equilibrium cell potential \cite{bard2000electrochemical,newman2012electrochemical} and also neglects higher-order effects. These effects can become important in experimental systems \cite{wang2022long, wang2022visualization, rath2021ph} where the voltage amplitude or the steady bias in the applied potential impacts the reaction rate, boundary conditions, and leading order solution. Therefore, the analysis presented here will need to be adjusted to incorporate these effects. We outline the modifications required to incorporate these effects in section \ref{sec: limitation}. 
\label{sec: numerics}
\subsection{Leading Order}
The leading-order limit refers to the condition of no applied potential, and thus the leading-order solutions are set by the initial conditions 
\begin{subequations}
\label{Eq: zero_order}
\begin{equation}
    \phi^{(0)}(x, t) = 0,
\end{equation}
\begin{equation}
    c_i^{(0)} (x,t) = c_{i0}.
\end{equation}
\end{subequations}
One can also verify that the solution above satisfies the governing equations and boundary conditions at the leading order.
\subsection{First Order}
 At the first order, Eq. (\ref{eq: nondimpoisson}) takes the form
 \begin{subequations}
\begin{equation}
    \frac{\partial^2 \phi^{(1)}}{\partial x^2} = -\kappa^2 \rho_e^{(1)}.
    \label{eq: nondimfirstorderpoisson}
\end{equation}
For the $i^{\textrm{th}}$ ion, Eq. (\ref{eq: nondimNP}) reduces to
\begin{equation}
\label{eq: nondimNPfirstorder}
    \frac{1}{D_i}\frac{\partial c_i^{(1)}}{\partial t} = \frac{\partial^2 c_i^{(1)}}{\partial x^2} + z_i  c_{i0} \frac{\partial^2 \phi^{(1)}}{\partial x^2}.
\end{equation}
\end{subequations}
In order to separate temporal and spatial variables, we consider solutions to Eqs. (\ref{eq: nondimfirstorderpoisson}) and (\ref{eq: nondimNPfirstorder}) of the forms $\phi^{(1)}(x,t) = \textrm{Im}\left[e^{i \omega t} \hat{\phi}^{(1)}(x) \right]$ and $c_i^{(1)}(x,t) = \textrm{Im}\left[e^{i \omega t} \hat{c}_i^{(1)}(x) \right]$. The governing equations with only spatial dependency become 
\begin{subequations}
\label{eq: nondimNPfirstorderspace}
\begin{eqnarray}
 \frac{d^2 \hat{\phi}^{(1)}}{d x^2} = -\kappa^2 \hat{\rho}_e^{(1)}, \\
 \frac{i \omega {\hat{c}_i^{(1)}}}{{D_i}} = \frac{d^2 \hat{c}_i^{(1)}}{d x^2} + z_i c_{i0} \frac{d^2 \hat{\phi}^{(1)}}{dx^2}. 
\label{eq: NP1hat}
 \label{eq: poisson1hat} 
\end{eqnarray}
\end{subequations}
Following a similar process for the boundary conditions listed in Eqs. (\ref{eq: nondimBCs}), we find the boundary conditions for Eqs. (\ref{eq: nondimNPfirstorderspace}) are 
\begin{subequations}
\label{eq: nondimBCshat}
\begin{eqnarray}
  \hat{\phi}^{(1)}\biggr{|}_{x = \pm 1} = \pm 1, \\  
- \left(\frac{d \hat{c_i}^{(1)}}{d x}  + z_i c_{i0}  \frac{d \hat{\phi}^{(1)}}{d x} \right)\biggr{|}_{x = \pm 1} = \frac{n_{i0}^{(1)}}{D_i}. 
\end{eqnarray}
\end{subequations}
Eqs. (\ref{eq: nondimNPfirstorderspace}) and (\ref{eq: nondimBCshat}) enable the determination of $\hat{c}_i^{(1)}$ and $\hat{\phi}^{(1)}$. Since the variables are periodic at this order, i.e., the average of $e^{i \omega t}$ is 0, neither an AREF nor an ARCF are observed. Thus, we examine the second order.
\par{} From the results of Eq. (\ref{eq: nondimNPfirstorderspace}), we write salt concentration or salt $\hat{s}^{(1)} = \sum_{i}{\hat{c}_i^{(1)}}$, charge density $\hat{\rho}_e^{(1)} = \sum_{i}{z_i \hat{c}_i^{(1)}}$, ionic strength $\hat{I}^{(1)} = \sum_{i}{z_i^2 \hat{c}_i^{(1)}}$, and electric field $\hat{E}^{(1)} = -\frac{d \hat{\phi}^{(1)}}{d x}$. These variables are employed at the second order. 
\subsection{Second Order}
We time average the governing equations at the second order over one period of the applied potential such that Eq. 
(\ref{eq: nondimpoisson}) reads 
\begin{subequations}
\label{eq: nondimPNPsecondorder}
\begin{equation}
\label{eq: nondimpoissonsecondorder}
 \frac{d^2 \left< \phi^{(2)}\right>}{d x^2} = -\kappa^2 \left< \rho_e^{(2)}\right>.
 \end{equation}
We add Eqs. (\ref{eq: nondimNP}) for all ions and time average to get 
 \begin{equation}
\frac{d ^2 \left<s^{(2)}\right>}{d x^2} -  \frac{1}{2} \frac{d}{d x} \textrm{Re}\left( \hat{\rho}_e^{(1)} \bar{E}^{(1)}\right) = 0.
      \label{eq: nondimsalt}
\end{equation}
We multiply Eq. (\ref{eq: nondimNP}) by $z_i$, time average, and sum the equations to obtain 
 \begin{equation}
\frac{d ^2 \left<\rho_{e}^{(2)}\right>}{d x^2} + I_0 \frac{d^2 \left<\phi^{(2)}\right>}{d x^2} -  \frac{1}{2} \frac{d}{d x} \textrm{Re}\left( \hat{I}^{(1)} \bar{E}^{(1)}\right) = 0.
      \label{eq: nondimNPsecondorder}
\end{equation}
\end{subequations}
Variables with bar are complex conjugates of variables with hat, and $\left< \right>$ corresponds to time-averaged variables. Note that  Eqs. (\ref{eq: nondimPNPsecondorder}) are sufficient to determine the presence and forms of the AREF and ARCF.  Further, we determine that diffusivity has no explicit effect on the second order time-averaged results, though $D_i$ indirectly influences the first-order variables. \\ 
The boundary conditions given in Eqs. (\ref{eq: nondimBCs}) at the second order become 
\begin{subequations}
    \label{eq: nondimBCsSecondOrder}
    \begin{eqnarray}
           \left(\frac{d \left< s^{(2)} \right>}{d x}  - \frac{1}{2} \textrm{Re}\left( \hat{\rho}_e^{(1)} \bar{E}^{(1)} \right) \right)\biggr{|}_{x = \pm 1} = 0,  \label{eq: fluxsaltSecondOrder} \\
       \left(\frac{d \left< \rho_e^{(2)} \right>}{d x}  + I^{(0)}  \frac{d \left<\phi^{(2)}\right>}{d x} - \frac{1}{2} \textrm{Re}\left( \hat{I}^{(1)} \bar{E}^{(1)} \right) \right)\biggr{|}_{x = \pm 1} = 0,  \label{eq: fluxSecondOrder}\\
       \left. \left< \phi^{(2)} \right> \right|_{x = \pm 1} = 0 \label{eq: potsecondorder}.
    \end{eqnarray}
\end{subequations}
We integrate Eqs. (\ref{eq: nondimsalt}) and (\ref{eq: nondimNPsecondorder}) with boundary conditions in Eqs. (\ref{eq: fluxsaltSecondOrder}) and (\ref{eq: fluxSecondOrder}) to write
\begin{subequations}
\begin{eqnarray}
    \frac{d \left< s^{(2)} \right>}{d x}  - \frac{1}{2} \textrm{Re}\left( \hat{\rho}_e^{(1)} \bar{E}^{(1)} \right) = 0 \label{eq: secondordersaltfinal}, \\
        - \frac{1}{\kappa^2} \frac{d^3 \left<\phi^{(2)}\right>}{d x^3} + I_0  \frac{d \left<\phi^{(2)}\right>}{d x} -  \frac{1}{2} \textrm{Re}\left( \hat{I}^{(1)} \bar{E}^{(1)}\right) = 0, 
    \label{eq: secondorderfinalNP}
\end{eqnarray}
\end{subequations}
where we have also utilized Eq. (\ref{eq: nondimpoissonsecondorder}). Note that we define the ARCF as the salt gradient $\frac{d \left< s^{(2)} \right>}{dx}$. To integrate $\left< s^{(2)} \right>$ using Eq. (\ref{eq: secondordersaltfinal}), since the flux of salt is zero at both boundaries at the second order, $ \int_{-1}^1 \left<s^{(2)} dx \right> = 0$ can be used as a boundary condition. 
\par{} Similarly, Eq. (\ref{eq: secondorderfinalNP}) is a third-order equation in $\left< \phi^{(2)} \right>$, but we only have two boundary conditions in Eq. (\ref{eq: potsecondorder}).  To find the third boundary condition \cite{jarvey2022ion}, we note that since the flux of charge is zero at both boundaries at the second order, $ \int_{-1}^1 \left<\rho_e^{(2)} dx \right> = 0$. By substituting Eq. (\ref{eq: nondimpoissonsecondorder}) in the aforementioned condition, we obtain
\begin{equation}
\label{eq: addBCsecondorder}
    \left. \frac{d \left< \phi^{(2)} \right> }{d x} \right|_{x=1} -     \left. \frac{d \left< \phi^{(2)} \right> }{d x} \right|_{x=-1} =0. 
\end{equation}
Eq. (\ref{eq: secondorderfinalNP}) can thus be solved with Eqs. (\ref{eq: potsecondorder}) and (\ref{eq: addBCsecondorder}) as boundary conditions.  \\
Experimentally \cite{woehl2015bifurcation, bukosky_extreme_2019,bukosky2015simultaneous,saini2016influence,wang2022long,rath2021ph}, the relevant limit is thin EDLs, or $\kappa \gg 1$. In this limit, a singular perturbation is required where solutions are divided into the EDL regions and the region outside the EDLs \cite{jarvey2022ion, balu2022electrochemical, balu_thin_2021}. However, since AREFs occur outside the EDLs, one can simplify Eq. (\ref{eq: secondorderfinalNP}) in the limit $\kappa \gg 1$ to directly write
\begin{equation}
    \left< E^{(2)}\right> = -\frac{1}{2 I^{(0)}} \textrm{Re}\left( \hat{I}^{(1)} \bar{E}^{(1)} \right),
    \label{eq: analyticsAREFfinal}
\end{equation}
where $\left< E^{(2)}\right> = - \frac{d \left< \phi^{(2)} \right>}{dx}$. We note that Eq. (\ref{eq: analyticsAREFfinal}) is valid only outside the EDL regions, and thus directly predicts the AREF. Eq. (\ref{eq: analyticsAREFfinal}) highlights that the presence of an AREF is only dependent on the first-order ionic strength and the first-order electric field. Even further, it is known that as $\bar{E}^{(1)} \neq 0$ in the concentration boundary layer due to the asymmetric boundary conditions (see Eq. (\ref{eq: nondimBCshat}) and the discussion in section \ref{sec: binary}), $\left< E^{(2)}\right>$ will be nonzero when $\hat{I}^{(1)} \neq 0$.  Physically, this implies that an imbalance in ionic strength outside the EDL regions creates an AREF. This is a crucial physical insight that our analysis reveals. We would like to emphasize that this requirement is true for an arbitrary number of ions without any restriction on valences and diffusivities, albeit within the limits of thin double layers and small potentials. 
\par{} We provide a brief physical explanation regarding the requirement of $\hat{I}^{(1)} \neq 0$ to create an AREF. Physically, EDL charging and/or redox reactions produce electric currents which subsequently induce a net electric field in the regions outside the EDLs. Still, electroneutrality is required to hold in the regions excluding the EDLs. Therefore, outside the EDLs, the only possible charge flux is an electromigrative flux, which has a magnitude that is directly dependent on the local ionic strength. This induced imbalance in ionic strength produces an asymmetry in the local conductivity of the electrolyte, which results in a time-averaged charge flux. An AREF forms to balance this charge flux.
\par{} \begin{figure}[b!]
    \centering
    \includegraphics[width = 0.45 \textwidth]{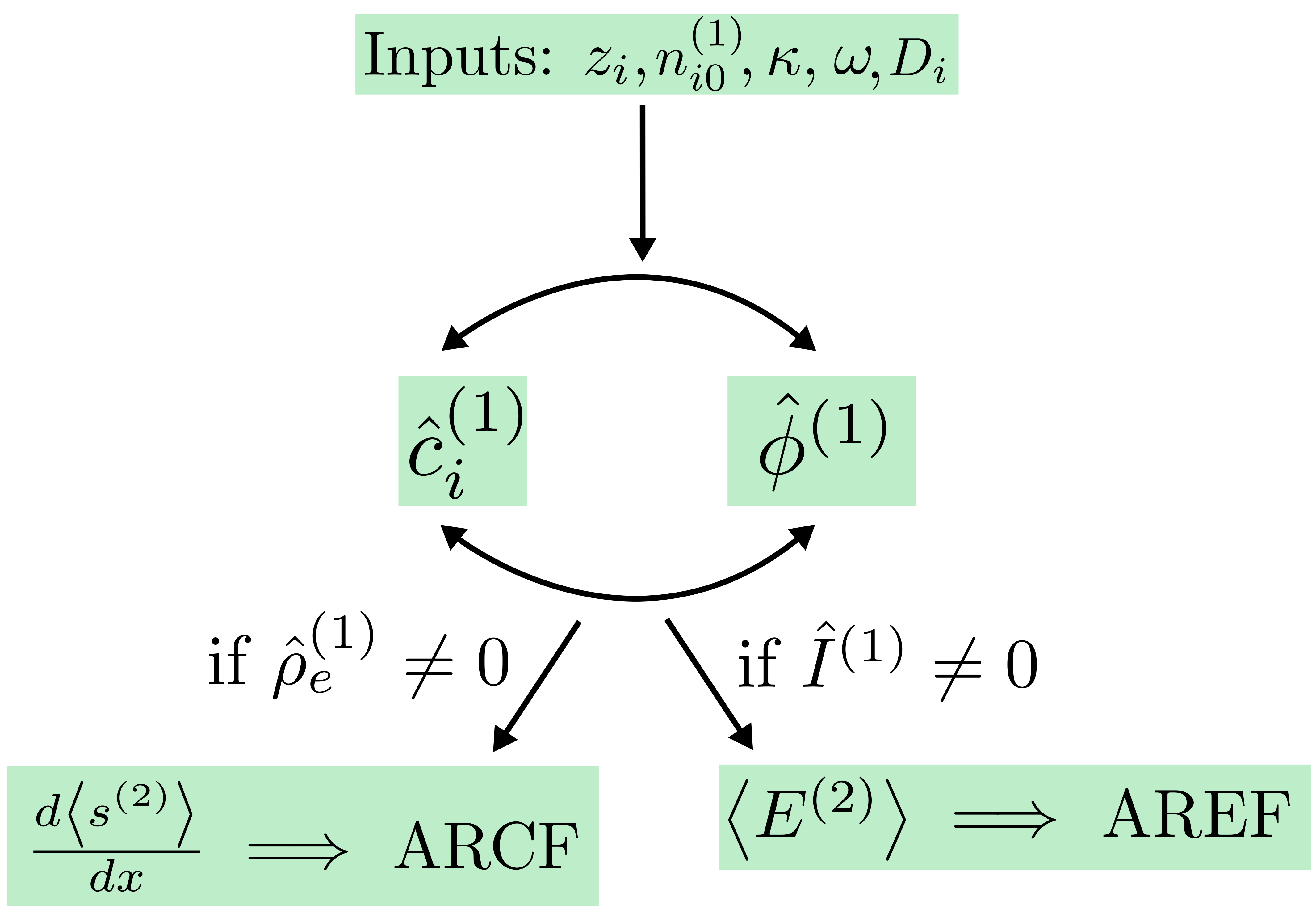}
    \caption{Methodology for calculating the AREF and ARCF analytically in low-potential and thin-double-layer limits for multi-ion electrolytes. The key finding is that a nonzero first-order ionic strength is a sufficient criterion to observe an AREF and a nonzero first-order charge density outside EDLs is a sufficient condition to observe an} ARCF.
    \label{Fig: Fig7}
\end{figure} Eq. (\ref{eq: secondordersaltfinal}) shows that there can be steady salt gradients at the second order. Since salt gradients give rise to diffusiophoresis \cite{prieve1984motion, anderson1989colloid, shin2016size, banerjee2019long, gupta2019diffusiophoretic, gupta2020diffusiophoresis, raj2023two}, the salt gradient or ARCF can simply be evaluated by utilizing Eq. (\ref{eq: secondordersaltfinal}). We emphasize that ARCFs are also a phenomenon that occurs outside EDLs and are induced by an imbalance in the first-order charge density; see Eq. (\ref{eq: secondorderfinalNP}). At this point, we note that diffusiophoretic mobility may vary for different
combinations of ions\cite{gupta2019diffusiophoretic}, and calculating a steady concentration field for each ion may be important in some scenarios. These calculations are straightforward, but we do not detail them for brevity.       
\par{} We summarize the procedure for estimating AREFs and ARCFs in the form of a flowchart in Fig. \ref{Fig: Fig7}. Ion valences, diffusivities, charge fluxes due to surface reactions, relative double-layer thickness, and frequency are the required inputs of the first-order solution. At the first order, the Poisson equation (\ref{eq: analyticspoissonfirstorder}) and the Nernst-Planck (NP) equation in charge (\ref{eq: analyticsrho1NP}) are solved together to obtain charge density and potential. They subsequently are used as inputs to determine the first-order ionic strength. Once the first-order charge density, potential (and by extension electric field), and ionic strength are determined, the AREF is evaluated using Eq. (\ref{eq: analyticsAREFfinal}). A nonzero $\hat{I}^{(1)}$ outside of the EDLs indicates a nonzero $\left< E^{(2)}\right>$, and thus indicates the presence of an AREF. Similarly, first-order charge density and potential are used to determine the ARCF using Eq. (\ref{eq: secondordersaltfinal}). A nonzero $\hat{\rho}_e^{(1)}$ outside the EDLs is a requirement for an ARCF.

\subsection{Numerical Solution Procedure}
\label{subsec: numericsprocedure}
\par{} We first solve Eqs. (\ref{eq: nondimNPfirstorderspace}) with boundary conditions in Eqs. (\ref{eq: nondimBCshat}) using the \textit{bvp4c} functionality in MATLAB.  $D_i$, $z_i$, $c_{i0}$, $n_{i0}^{(1)}$, $\kappa$ and $\omega$ are used as inputs to our code. \textit{bvp4c} deploys an adaptive grid meshing in the spatial dimension. We benchmark these numerical results with the results from Hashemi et al. \cite{hashemi2020perturbation} for a binary electrolyte with no reactions and asymmetric diffusivities. 
\par
\begin{figure}[t!]
    \centering
    \includegraphics[width = 0.45 \textwidth]{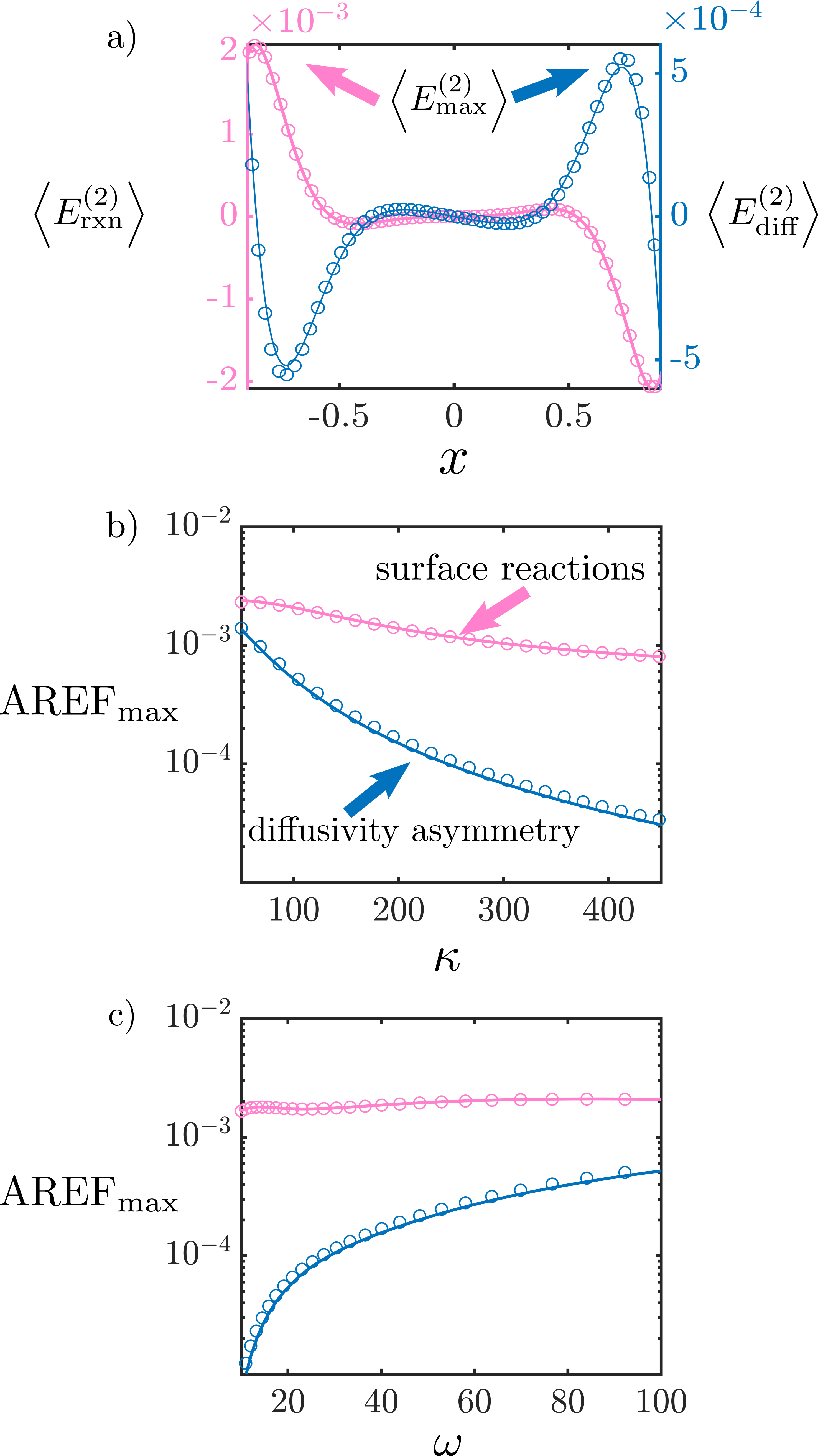}
    \caption{\textbf{Analysis of the} AREF for a binary electrolyte. }(a) $\langle E^{(2)}\rangle$ vs. $x$ for surface reactions (pink, left y-axis) and diffusivity asymmetry (blue, right y-axis) cases. Dependency of the maximum value of the AREF, i.e., $\left<E^{(2)}_{\textrm{max}} \right> = \textrm{AREF}_{\textrm{max}}$ on (b) $\kappa$ with $\omega = 100$ and (c) $\omega$ with $\kappa = 100$. The solid lines represent analytical solutions and the orbs are numerical calculations. The pink color represents surface reactions, and the blue color represents diffusivity asymmetry. All cases have ions with valences of $z_1 = 1$ and $z_2 = -1$, respectively. Surface reactions have $j^{(1)}_0 = -0.5$, $D_1 = 1$, and $D_2 = 1$, while diffusivity asymmetry has $j^{(1)}_0 = 0$, $D_1 = 2$, and $D_2 = 1$. 
    \label{Fig: Fig2}
\end{figure}

{}  Next, we also use \textit{bvp4c} to solve Eq. (\ref{eq: secondorderfinalNP}) with boundary conditions set by Eqs. (\ref{eq: potsecondorder}) and (\ref{eq: addBCsecondorder}). These equations require the additional input of $\hat{I}^{(1)} \bar{E}^{(1)}$, which we calculate from the first-order equations. Similarly, we solve Eq. (\ref{eq: secondordersaltfinal}) with the salt conservation boundary condition. This calculation requires the input of $\hat{\rho}_e^{(1)} \bar{E}^{(1)}$, which we obtain from the first-order equations. We  benchmark our results with the binary electrolyte results from Hashemi et al.\cite{hashemi2020perturbation}. Comparisons between numerical and analytical results are shown in Figs. \ref{Fig: Fig2} and \ref{Fig: Fig4}, and our adaptation of the analytics from Hashemi et al. \cite{hashemi2020perturbation} are given in the Appendix.
\section{Analytical Solution for Symmetric Diffusivites and Thin Double Layers}
\label{sec: analytics}
To make analytical progress, we assume equal diffusivities of all ions, i.e.,  $D_i = 1$, and the thin-double-layer limit, i.e., $\kappa \gg 1$. Eq. (\ref{eq: nondimNP}) at first order in $\phi_D$ for the  variables $\hat{\rho}_e^{(1)}$, $\hat{I}^{(1)}$, and  $\hat{\phi}^{(1)}$ becomes 
\begin{subequations}
\begin{eqnarray}
 - \frac{d^2 \hat{\phi}^{(1)}}{d x^2} = \kappa^2 \hat{\rho}_e^{(1)}, \label{eq: analyticspoissonfirstorder} \\
i \omega {\hat{\rho}_e^{(1)}}= \frac{d^2 \hat{\rho}_e^{(1)}}{d x^2} + I^{(0)} \frac{d^2 \hat{\phi}^{(1)}}{d x^2},
\label{eq: analyticsrho1NP} \\
i \omega {\hat{I}^{(1)}}  = \frac{d^2 \hat{I}^{(1)}}{d x^2} + \sum_{i}{z_i^3 c_{i0}} \frac{d^2 \hat{\phi}^{(1)}}{d x^2}, \label{eq: analyticsNPI1} 
\end{eqnarray}
\end{subequations} 
The corresponding boundary conditions for these equations are given by
\begin{subequations}
\begin{eqnarray}
  \hat{\phi}^{(1)}\biggr{|}_{x = \pm 1} = \pm 1, \label{eq: phiBCsanalytics1} \\
 -\left(\frac{d \hat{\rho}_e^{(1)}}{d x} + I^{(0)}  \frac{d \hat{\phi}^{(1)}}{d x} \right)\biggr{|}_{x = \pm 1} = \sum_{i}{z_i n_{i0}^{(1)}}, \label{eq: rhoBCsanalytics1} \\
  -\left(\frac{d \hat{I}^{(1)}}{d x} + \sum_{i}{z_i^3 c_{i0}}  \frac{d \hat{\phi}^{(1)}}{d x} \right)\biggr{|}_{x = \pm 1} = \sum_{i}{z_i^2 n_{i0}^{(1)}}. \label{eq: IBCsanalytics1} 
\end{eqnarray}
\end{subequations}
We require asymmetric solutions to these equations since $\hat{\phi}^{(1)}$ has asymmetric boundary conditions. We begin by finding an expression for $\hat{\rho}_e^{(1)}$. As a shorthand notation, let $\lambda_1 = \sqrt{i \omega}$ and $\lambda_2 = \sqrt{I^{(0)} \kappa^2 + i \omega}$. Then, by substituting Eq. (\ref{eq: analyticspoissonfirstorder}) into Eq. (\ref{eq: analyticsrho1NP}) and integrating, we determine
\begin{equation}
\label{eq: analyticsrho1eqn1} 
    \hat{\rho}_e^{(1)} = B \sinh{\left( \lambda_2 x \right)},
\end{equation}
where $B$ is an unknown constant. Eq. (\ref{eq: analyticsrho1eqn1}) is utilized  in Eq. (\ref{eq: analyticspoissonfirstorder}) and integrated twice to find
\begin{subequations}
\begin{equation}
\label{eq: analyticsphi1eqn1}
    \hat{\phi}^{(1)} = x + B \frac{\kappa^2}{\lambda_2^2} \left[ x \sinh{\left( \lambda_2 \right)} - \sinh{\left( \lambda_2 x\right)} \right],
\end{equation}
where we have already employed the boundary conditions in Eq. (\ref{eq: phiBCsanalytics1}). With functional forms for both $\hat{\phi}^{(1)}$ and $\hat{\rho}_e^{(1)}$, we use Eq. (\ref{eq: rhoBCsanalytics1}) to show 
\begin{equation}
    B = -\frac{\lambda_2^2 \left[j^{(1)}_0 + I^{(0)} \right]}{I^{(0)}  \kappa^2 \sinh{\left( \lambda_2 \right)} + \lambda_1^2 \lambda_2 \cosh{\left( \lambda_2 \right)}},
    \label{eq: B}
\end{equation}
\end{subequations}
where $j^{(1)}_0 = \sum_{i}{z_i n_{i0}^{(1)}}$. With the full forms of $\hat{\phi}^{(1)}$ and $\hat{\rho}_e^{(1)}$ determined, we solve for $\hat{I}^{(1)}$. Based upon Eq. (\ref{eq: analyticsNPI1}), we observe that we will require a particular and a homogeneous solution to determine $\hat{I}^{(1)}$ due to the inhomogeneity brought about by the electromigrative term. Rewriting Eq. (\ref{eq: analyticspoissonfirstorder}) in the form of an operator on the left side, we show
\begin{equation}
    \label{eq: analyticsI1NPrearrange}
    \left( \frac{d^2}{d x^2} - \lambda_1^2 \right) 
    \hat{I}^{(1)} = \sum_{i}{z_i^3 c_{i0}} \kappa^2 \hat{\rho}_e^{(1)}.
\end{equation}
Note that from Eqs. (\ref{eq: analyticspoissonfirstorder}) and (\ref{eq: analyticsrho1NP}), $\left(\frac{d^2}{d x^2} - \lambda_2^2 \right) \hat{\rho}_e^{(1)} = 0$. This means that we can apply the operator $\left( \frac{d^2}{d x^2} - \lambda_2^2 \right)$ to both sides of Eq. (\ref{eq: analyticsI1NPrearrange}) to reach a homogeneous equation. Taking only the asymmetric solutions for $\hat{I}^{(1)}$, we arrive at
\begin{subequations}
\begin{equation}
    \label{eq: analyticsI1noBCsyet}
    \hat{I}^{(1)} = F \sinh{\left( \lambda_1 x \right)} + G \sinh{\left( \lambda_2 x \right)},
\end{equation}
where $F$ and $G$ are constants that need to be determined. We take $G$ such that it cancels out the inhomogeneity in Eq. (\ref{eq: analyticsI1NPrearrange}), which results in
\begin{equation}
    G = \frac{\sum_{i}{z_i^3 c_{i0}}}{I^{(0)}} B.
\end{equation}
Next, we apply the boundary conditions in Eq. (\ref{eq: IBCsanalytics1}) and determine
\begin{equation}
    F = -\frac{\sum_{i}{z_i^2 n_{i0}^{(1)}} +  \left(\sum_{i}{z_i^3 c_{i0}} \right) j^{(1)}_0 / I^{(0)}}{\lambda_1 \cosh{\lambda_1}}.
    \label{eq: F}
\end{equation}
\end{subequations}
With $\hat{\rho}_e^{(1)}$ and $\hat{I}^{(1)}$ fully determined,  we invoke Eq. (\ref{eq: analyticsAREFfinal}) to obtain an expression for the AREF. We also invoke Eq. (\ref{eq: secondordersaltfinal}) to calculate the ARCF.
\par{} The analytical expressions obtained shed light on the physics of the EDLs and concentration boundary layers. In the experimentally relevant limit of thin double layers relative to the length of the concentration boundary layer, i.e., $\omega/(I^{(0)}\kappa^2)\ll 1$, the charge density obtained from Eq. (\ref{eq: analyticsrho1eqn1}) is entirely screened over the EDL regions. However, there are salt dynamics introduced by the surface reactions, resulting in an ionic strength imbalance over the concentration BL regions, as seen from the homogeneous solution in Eq. (\ref{eq: analyticsI1noBCsyet}). AREFs, therefore, result from the simultaneous presence of first-order electric field and ionic strength. Local maxima of the components of ionic strength are found over the concentration boundary layer dimensionless length scale $\omega^{-1/2}$; see Fig. \ref{Fig: Fig1}(b). Simultaneously, surface reactions, an AC field, or both effects can produce a residual electric field in the concentration boundary layers. In fact, it can be shown by integrating Eqs. (\ref{eq: analyticspoissonfirstorder}) and (\ref{eq: analyticsrho1NP}) over the EDLs that in order for the combination of diffusive and electromigrative fluxes to match the surface  charge flux and the rate of change of accumulated charge in the EDLs, there must be a homogeneous residual electric field in the concentration BLs. As seen from Eq. (\ref{eq: analyticsphi1eqn1}), this residual field is given by $\hat{E}^{(1)} = -1 - \frac{B \kappa^2\sinh(\lambda_2)}{\lambda_2^2}\neq 0$, which leads to the conclusion that only a nonzero $\hat{I}^{(1)}$ is requisite to lead to an AREF. Additionally, we note that the charge density is fully screened over the EDLs in this scenario, meaning no ARCF will develop (see also section \ref{sec: arcf}).
\begin{figure*}[t!]
    \centering
    \includegraphics[width = 0.9 \textwidth]{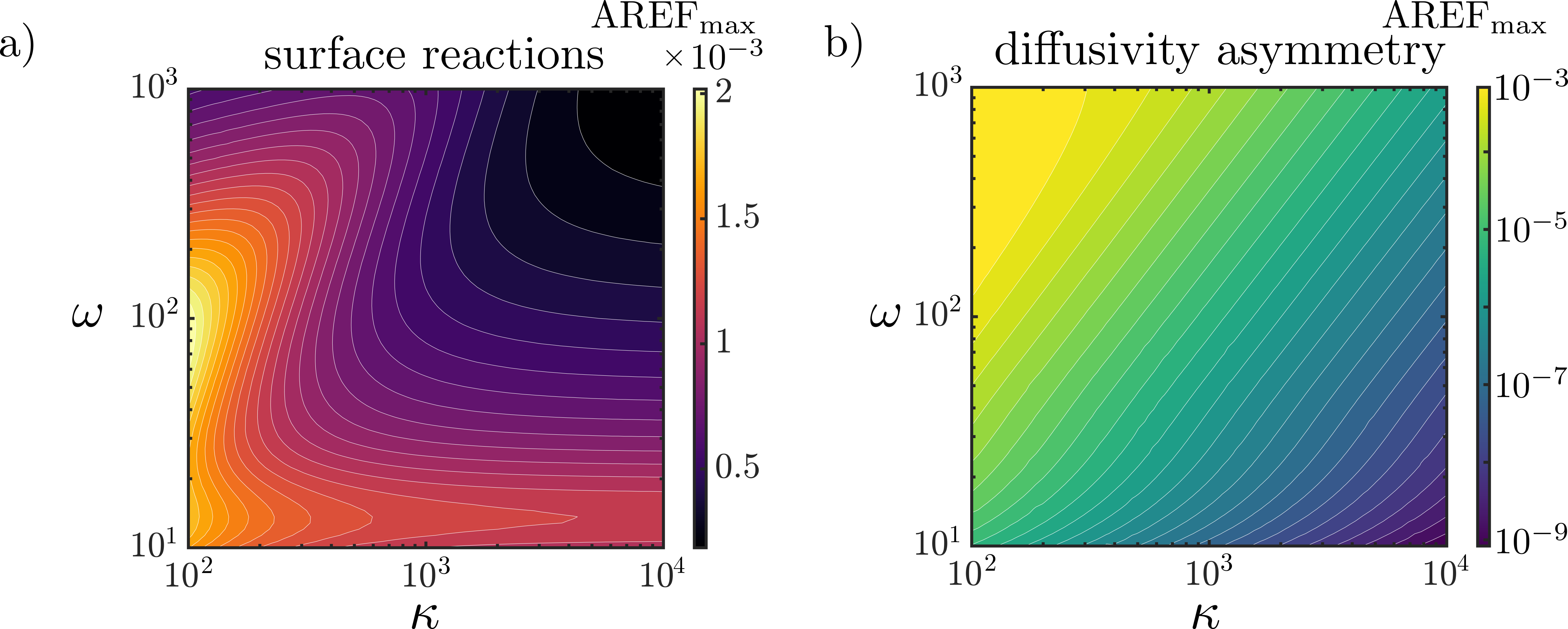}
    \caption{\textbf{Contour plot of the} maximum AREF for binary electrolyte for (a) surface reactions} only and (b) diffusivity asymmetry only. Panel (a) is simulated with $j^{(1)}_0 = -0.5$, $D_1 = 1$, and $D_2 = 1$, while panel (b) is simulated with $D_1 = 2$, $j^{(1)}_0 = 0$, and $D_2 = 1$. Both cases have two ions with valences of $z_1 = 1$ and $z_2 = -1$, respectively. Variation of AREF$_\textrm{max}$ for the surface reactions case alone is less sensitive to changes in $\kappa$ and $\omega$ than the diffusivity asymmetry case  alone. 
    \label{Fig: Fig3}
\end{figure*}
\section{Results and Discussion}
\label{sec: results}
\subsection{Asymmetric Rectified Electric Fields (AREFs)}
\subsubsection{Binary Electrolyte}
\label{sec: binary}
We first analyze the formation of an AREF due to surface reactions and compare it with the previously known requirement of diffusivity asymmetry to produce AREFs \cite{hashemi_oscillating_2018, hashemi2020perturbation, balu_thin_2021}.  We focus on the limit of  $\kappa \gg 1$ such that Eq. (\ref{eq: analyticsAREFfinal}) is valid. We begin by comparing the two mechanisms for a monovalent binary electrolyte. Fig. \ref{Fig: Fig2} displays a comparison between numerical (orbs) and analytical (solid lines) results between the two mechanisms for AREF formation, i.e. diffusivity asymmetry (blue) and surface reactions (pink). Numerical results are calculated as per the procedure outlined in section \ref{subsec: numericsprocedure}. Analytical results are determined by the approach described in section \ref{sec: analytics}. Results by Hashemi et al. \cite{hashemi2020perturbation}  are employed for the diffusivity asymmetry case to obtain first-order results which are subsequently utilized in Eq. (\ref{eq: analyticsAREFfinal}) to obtain the AREF; see the Appendix for details. The surface reaction case is presented for $j^{(1)}_0 = -0.5$ and $D_1=D_2=1$, while the diffusivity asymmetry case is presented for $j^{(1)}_0 = 0$, $D_1 = 2$ and $D_2 = 1$.  We ignore the spatial regions located between $x = -1$ to $x = -0.9$ and $x = 0.9$ to $x = 1$ to focus on the regions outside the EDLs. As evident from Fig. \ref{Fig: Fig2}, we obtain excellent quantitative agreement between analytical and numerical results in all scenarios and for either mechanism. 
\par{} First, we note that an AREF is present even with symmetric diffusivities due to the presence of surface reactions, indicating a wider parameter space that can give rise to AREFs than was previously anticipated.  We observe that both the shape and the magnitude of the AREF are different for the two formation mechanisms employed; see Fig. \ref{Fig: Fig2}(a). Specifically, for the parameters chosen, reaction-driven  AREFs display a maximum near the left electrode, while AREFs produced by diffusivity asymmetry possess a maximum near the right electrode. The magnitude of the AREF maxima with surface reactions is greater than that of the AREF maxima with asymmetric diffusivities by roughly a factor of 5.  
\par{} We discuss the dependency of the maximum value of the AREF, denoted here by $\textrm{AREF}_{\textrm{max}}$, with $\kappa$ in Fig. \ref{Fig: Fig2}(b). For both mechanisms, $\textrm{AREF}_{\textrm{max}}$ decreases in magnitude as $\kappa$ increases. However, the decrease observed with surface reactions is significantly lower than the decrease observed with diffusivity asymmetry; we find that AREFs with diffusivity asymmetry decay as $\kappa^{-2}$, consistent with Balu and Khair \cite{balu_thin_2021}. Physically, AREF formation due to diffusivity asymmetry is driven by the currents arising from the EDLs. An increase in $\kappa$ reduces the volume of charge (and the current) in the EDLs. Therefore, AREF$_{\textrm{max}}$ also decreases. On the other hand, the current produced by surface reactions is not directly related to $\kappa$, leading to a  weaker dependence of $\textrm{AREF}_{\textrm{max}}$ on $\kappa$. The variation of $\textrm{AREF}_{\textrm{max}}$ with  $\omega$ is given in Fig. \ref{Fig: Fig2}(c). The surface reactions $\textrm{AREF}_{\textrm{max}}$ is insensitive to the change in $\omega$, while the diffusivity asymmetry $\textrm{AREF}_{\textrm{max}}$ increases with an increase in $\omega$. For $\omega = 100$, $\textrm{AREF}_{\textrm{max}}$ with reactions is still greater than with $D_i$ asymmetry by roughly a factor of 5. 
\par{}  The results outlined in Fig. \ref{Fig: Fig2} (b,c) demonstrate that if surface reactions are present, AREFs could be stronger than the AREFs created by diffusivity asymmetry alone, at least for the parameters chosen. To better understand the dependencies of $\textrm{AREF}_{\textrm{max}}$ on $\kappa$ and $\omega$, we employ numerical results to expand our parameter sweep to $\kappa = 10^2-10^4$ and $\omega = 10-10^3$ in Fig. \ref{Fig: Fig3}. Fig. \ref{Fig: Fig3}(a) shows the results for AREFs driven by surface reactions. The largest values of $\textrm{AREF}_{\textrm{max}}$ occur around $\kappa = 100$ and $\omega = 100$. Additionally, the different contours shown are all within an order of magnitude of one another, indicating that $\textrm{AREF}_{\textrm{max}}$ is weakly dependent on $\kappa$ and $\omega$. This weak dependency for a constant reactive charge flux can be understood through the dependency of first-order ionic strength. A nonzero $\hat{I}^{(1)}$ with surface reactions is impacted strongly by the boundary conditions $n_{i0}^{(1)}$ and weakly due to the effects of $\kappa$ and $\omega$, as shown analytically in Eqs. (\ref{eq: B}) and (\ref{eq: F}). Furthermore, our analysis shows that $\bar{E}^{(1)}$ also has a strong dependency on the surface reactive flux and a weak dependency on the frequency of the applied field. These coupled dependencies directly lead to the non-monotonic behavior observed in Fig. \ref{Fig: Fig3}(a), and also explain why $\textrm{AREF}_{\textrm{max}}$ values are within an order of magnitude of one another for a wide range of $\kappa$ and $\omega$ values.
\par{} In contrast, Fig. \ref{Fig: Fig3}(b) shows the contours of $\textrm{AREF}_{\textrm{max}}$ with $\kappa$ and $\omega$ values for AREFs caused by diffusivity asymmetry. Here, a monotonic increase in AREF$_{\textrm{max}}$ with an increase in $\omega$ and a decrease in $\kappa$ are observed. Unlike the surface reactions case, the differences in $\textrm{AREF}_{\textrm{max}}$ are on the scale of several orders of magnitude between different contours. This indicates a strong dependency of the AREF on both $\kappa$ and $\omega$. 
\par{} The results outlined in Figs. \ref{Fig: Fig2} and \ref{Fig: Fig3} emphasize that the behavior of AREFs with surface reactions is different from the behavior of AREFs with asymmetric diffusivities alone. Specifically, we find that the magnitude of an AREF tends to be larger with surface reactions for the range of parameters explored. In fact, the maximum values are also insensitive to $\kappa$ and $\omega$, meaning surface reactions are an important mechanism to tune AREFs. Next, we discuss AREFs in the presence of more than two ions.  
\begin{figure*}[t!]
    \centering
    \includegraphics[width = 0.9 \textwidth]{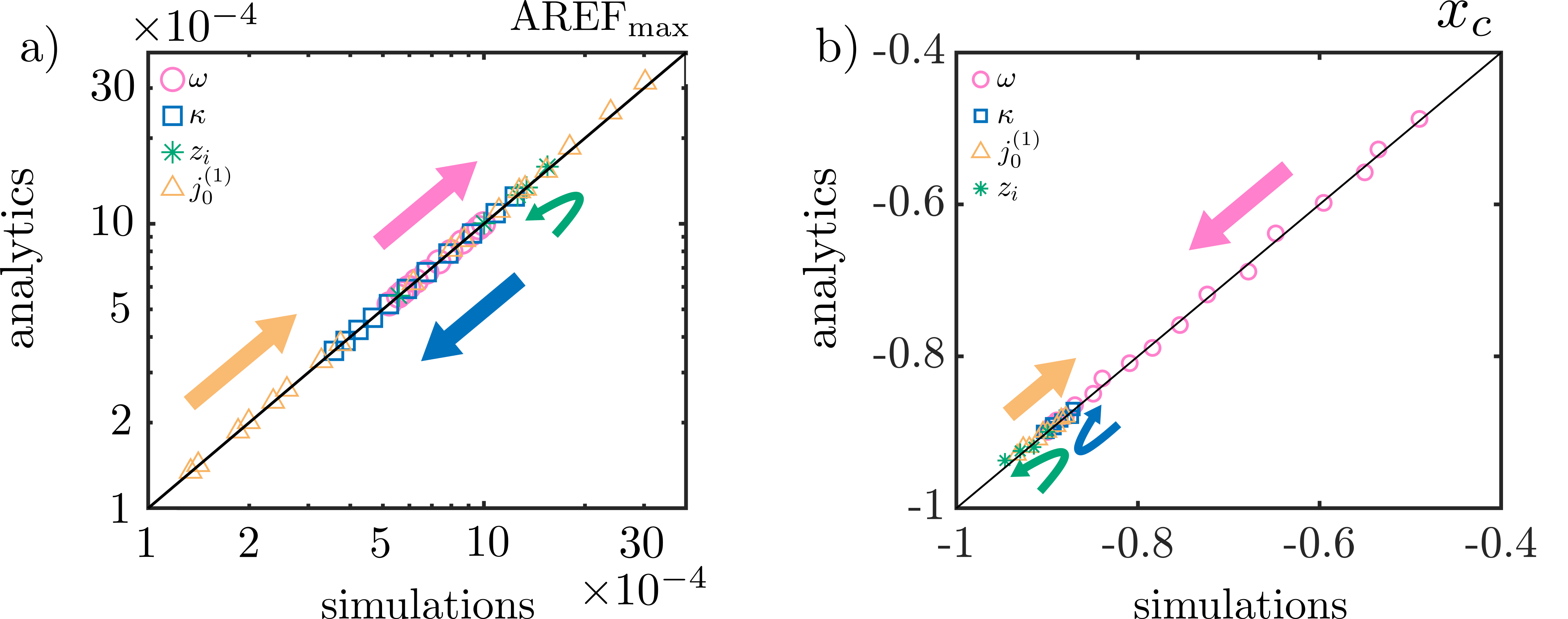}
    \caption{\textbf{Comparison of data} for analytical calculations and simulations for an electrolyte solution with three ions} for (a) the maximum value of AREF, i.e., AREF$_{\textrm{max}}$ and (b) the location of the maximum AREF near left electrode, i.e., $x_{\textrm{c}}$.  In both panels, $j^{(1)}_0 = -0.01\ \textrm{to}\ -0.85$ (orange), $\kappa = 50\ \textrm{to}\ 400$ (blue), $\omega = 10\ \textrm{to}\ 100$ (pink), and both $z_1 = -3\ \textrm{to}\ 3$ and $z_2 = -2, -3$ (green). When not varied, $\kappa = 100$, $\omega = 100$, $j^{(1)}_0 = -0.5$, $z_1 = 1$, $z_2 = -1$, and $z_3 = 1$. In all cases, we find that the results from analytics and numerical simulations collapse onto the diagonal, indicating strong agreement between the results of the two methods.
    \label{Fig: Fig4}
\end{figure*}
\subsubsection{Three Ions}
\label{subsec: threeion}
To move beyond the comparisons outlined in the prior subsection, we investigate the parameter dependencies of AREF$_{\textrm{max}}$ and its location, denoted here by $x_{c}$, for a solution with three ions. We note that $x_c$ is a crucial physical parameter that has been previously observed in experiments \cite{bukosky2015simultaneous, rath2021ph, wang2022long, bukosky_extreme_2019, woehl_electrolyte-dependent_2014, saini2016influence}.  
\par{} First, we compare the values of AREF$_{\textrm{max}}$ obtained from both numerical simulations and analytical calculations for a three-ion cell; see Fig. \ref{Fig: Fig4}(a). To focus on the effect of surface reactions, we keep diffusivities constant, or $D_1=D_2=D_3=1$. We perform a comprehensive sweep of parameters including $\omega$, $\kappa$, $z_i$ and $j_0^{(1)}$. Excellent quantitative agreement is observed between simulations and analytical solutions obtained over the entire space of parameters. Since the dependency of AREF$_{\textrm{max}}$ on $\kappa$ and $\omega$ was discussed previously, we focus this discussion on the dependency of AREF$_{\textrm{max}}$ on the other listed parameters. We find that $\textrm{AREF}_{\textrm{max}}$ increases with an increase in $j^{(1)}_0$. This trend is expected since increasing the the value $j^{(1)}_0$ leads to a larger $\hat{I}^{(1)}$, which  dictates the strength of the AREF. The changes in valence, in contrast, result in a non-monotonic behavior. The non-monotonic behavior of $z_i$ can be explained from the trend of $\kappa$. We recall that $\kappa^{-1}$ is not the true Debye length, but instead a measure of Debye length in the system. This choice was made out of mathematical convenience; see section \ref{sec: nondimensional}. As such, the green arrow  in Fig. \ref{Fig: Fig4}(a) starts at  $z_1=-3$ and ends at $z_1=3$. Therefore, the true Debye length first increases and then decreases. In effect, the trend should follow a decrease in $\kappa$ (i.e., opposite to blue arrow) and then an increase in $\kappa$ (i.e., in the direction of the blue arrow). This is consistent with the observed behavior of change in ionic valence. 
\par
\begin{figure}[t!]
    \centering
    \includegraphics[width = 0.45 \textwidth]{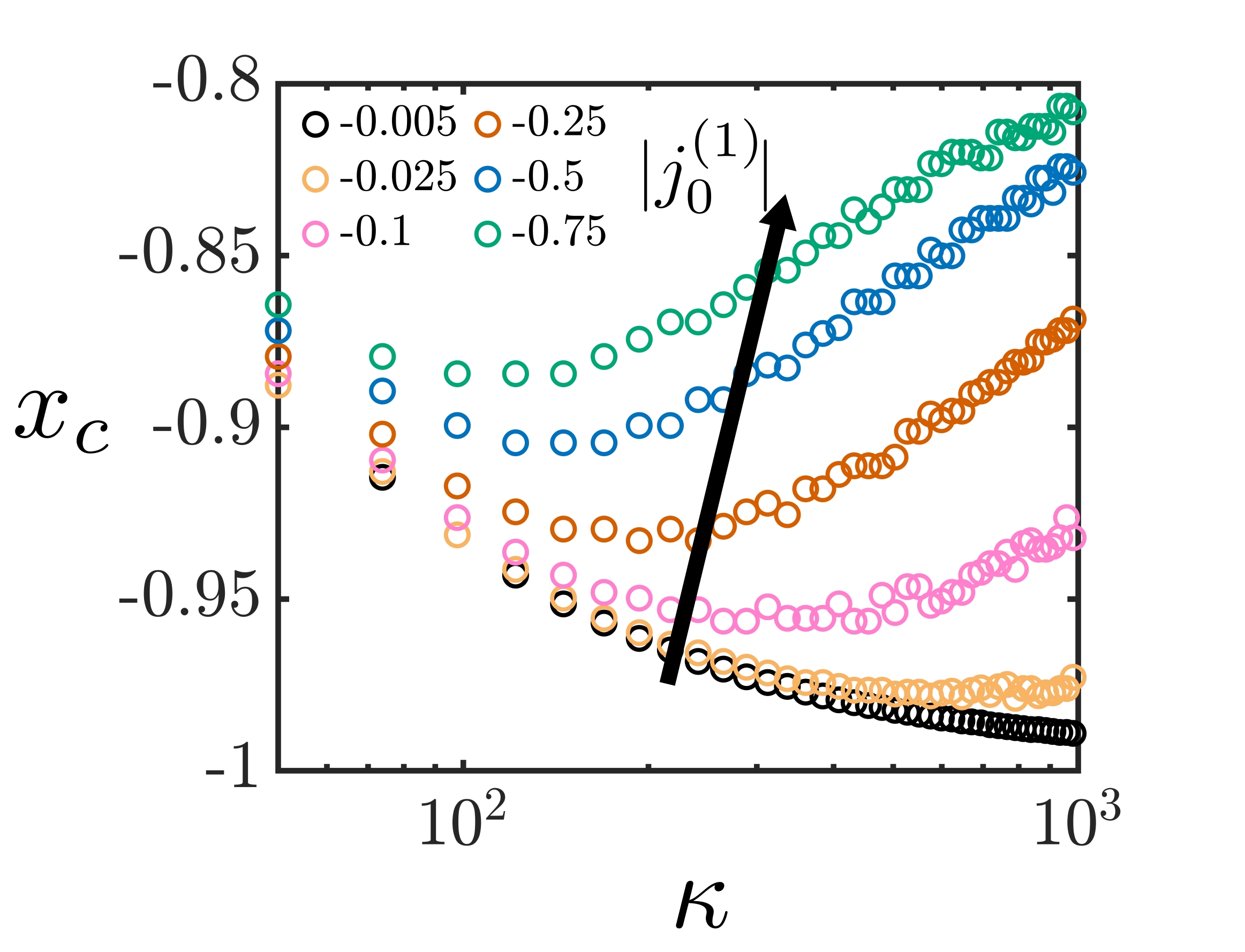}
    \caption{\textbf{Dependency of the maximum AREF location on $\kappa$ and $j^{(1)}_0$ for a three-ion electrolyte.} $\kappa$ vs. $x_c$ for $j^{(1)}_0 = -0.005, -0.025, -0.01, -0.25, -0.5, -0.75$. For all cases, $\omega = 100$, $z_1 = 1$, $z_2 = -1$, $z_3 = 1$, and $D_1 = D_2 = D_3 = 1$. }
    \label{Fig: Fig5}
\end{figure}

{} We now focus on $x_c$ closest to the left electrode; see Fig. \ref{Fig: Fig4}(b).  We observe strong quantitative agreement between simulations and analytical calculations. An increase in $j^{(1)}_0$ magnitude moves the location of the maximum further from the electrode. In contrast, an increase in $\omega$ moves the location of the maximum closer to that of the electrode. Surprisingly, the location of the maximum yields a non-monotonic behavior with $\kappa$. We investigate this in more detail in Fig. \ref{Fig: Fig5}.
\begin{figure*}[t!]
    \centering
    \includegraphics[width = 0.8 \textwidth]{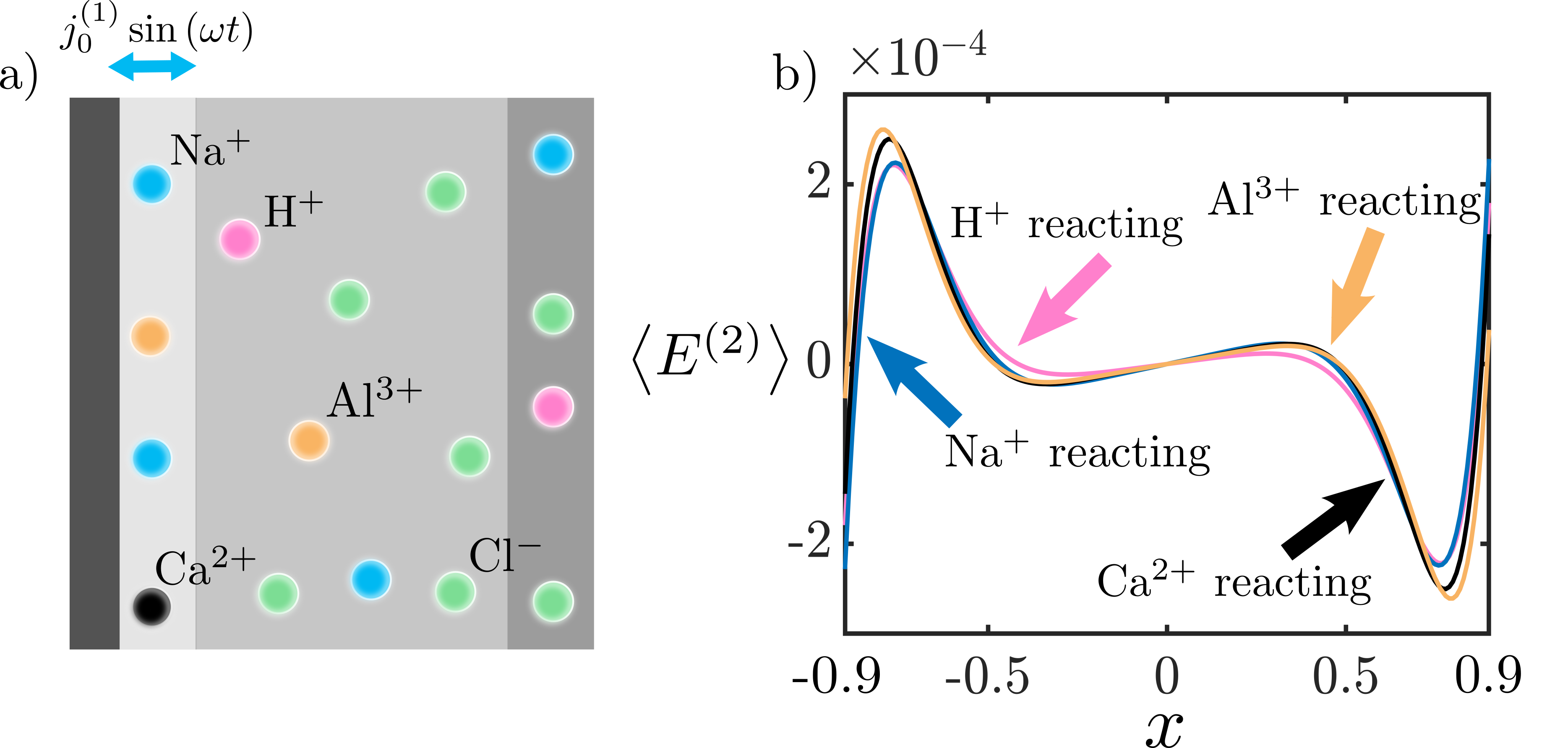}
    \caption{\textbf{Impact of the reactive ion on the} AREF in an electrolyte solution with 5 ions.} (a) Schematic of the model problem with H$^+$, Na$^+$, Ca$^{2+}$, Al$^{3+}$, and Cl$^-$, and a reactive charge flux $j^{(1)}_0$. (b) $\left< E^{(2)} \right>$ vs. $x$ with H$^+$ (pink), Na$^+$ (blue), Ca$^{2+}$ (black), and Al$^{3+}$ (orange) reacting, respectively. $j^{(1)}_0 = -0.5$, $\kappa = 100 $, $\omega = 100$, $c_{\textrm{Na}0} = 1$, $c_{\textrm{H}0} = 1$, $c_{\textrm{Ca}0} = 1$, $c_{\textrm{Al}0} = 1$, and $c_{\textrm{Cl}0} = 7$ in all cases. Results are obtained through numerical simulations.
    \label{Fig: Fig6}
\end{figure*}
To probe this non-monotonic behavior, Fig. \ref{Fig: Fig5} shows the location of the maximum with $\kappa$ for six different values of $j^{(1)}_0$ ranging from -0.005 to -0.75. For small $j^{(1)}_0$ values, we observe that the location moves towards the electrode with an increase in $\kappa$. However, as $j^{(1)}_0$ grows, this behavior becomes more non-monotonic, suggesting that competing effects are present in the system. In fact, when $\kappa$ values are small, the different curves of $j^{(1)}_0$ appear to converge.  We believe the non-monotonic behavior present is due to $\kappa$ driving movement of $x_c$ towards the electrode whereas  $j^{(1)}_0$ is driving $x_c$ away from the electrode. 
\par{} The results described in Fig. \ref{Fig: Fig4} demonstrate that the analytical procedure outlined in this manuscript is able to uncover the complex dependencies of AREFs on system parameters even for more than 2 ions. However, our analytical methodology is limited to the scenario of equal diffusivities. Asymmetric diffusivities can be included in the analytical formulation, but the relevant physics are not clearly visible with asymmetric diffusivities. As such, we solve the case with both asymmetric diffusivities and surface reactions numerically for convenience. With this in mind, we next focus on a 5-ion case with both surface reactions and diffusivity asymmetries. 
\subsubsection{Five-Ion Case}

\label{subsec: fiveions}
We investigate a five-ion problem ($\textrm{Na}^{+}$, $\textrm{H}^{+}$, $\textrm{Ca}^{2+}$, $\textrm{Al}^{3+}$, and $\textrm{Cl}^{-}$, where the diffusivity values for each ion are taken from literature \cite{haynes2016crc}) such that only one of the cations is reactive; a schematic of the problem is provided in Fig. \ref{Fig: Fig6}(a).  We would like to emphasize that the setup described here is hypothetical and these surface reactions may not be necessarily observable in a real electrochemical cell. The intent of this exercise is to study the simultaneous impact of diffusivity asymmetry and surface reactions, such as the ones present in experiments \cite{wang2022long, wang2022visualization, rath2021ph}. 
\par{} We focus on a constant charge flux magnitude due to surface reactions $j^{(1)}_0$. The resulting AREFs given in Fig. \ref{Fig: Fig6}(b) show the cases where each of the cations is reactive with $j_0^{(1)} = -0.5$. We note that the maximum AREF values are smaller that previously disscussed scenarios due to the inverse relationship between AREF$_{\textrm{max}}$ and $I^{(0)}$; see Eq. (\ref{eq: analyticsAREFfinal}). 
\par{} We find that the peak AREF location and magnitude is weakly dependent upon which ion is reacting. Even though the diffusivity of the $\textrm{H}^{\textrm{+}}$ ion is  larger than the other reacting ions by approximately one order of magnitude, no appreciable change in the AREF is observed. This indicates that if ionic diffusivities are on the same order of magnitude and $j_0^{(1)}=O(1)$, estimating AREFs by assuming equal diffusivities could serve as a good starting point. This underscores the utility of the analytical procedure outlined in section \ref{sec: analytics}.  

\begin{figure*}[t!]
    \centering
    \includegraphics[width = 0.8 \textwidth]{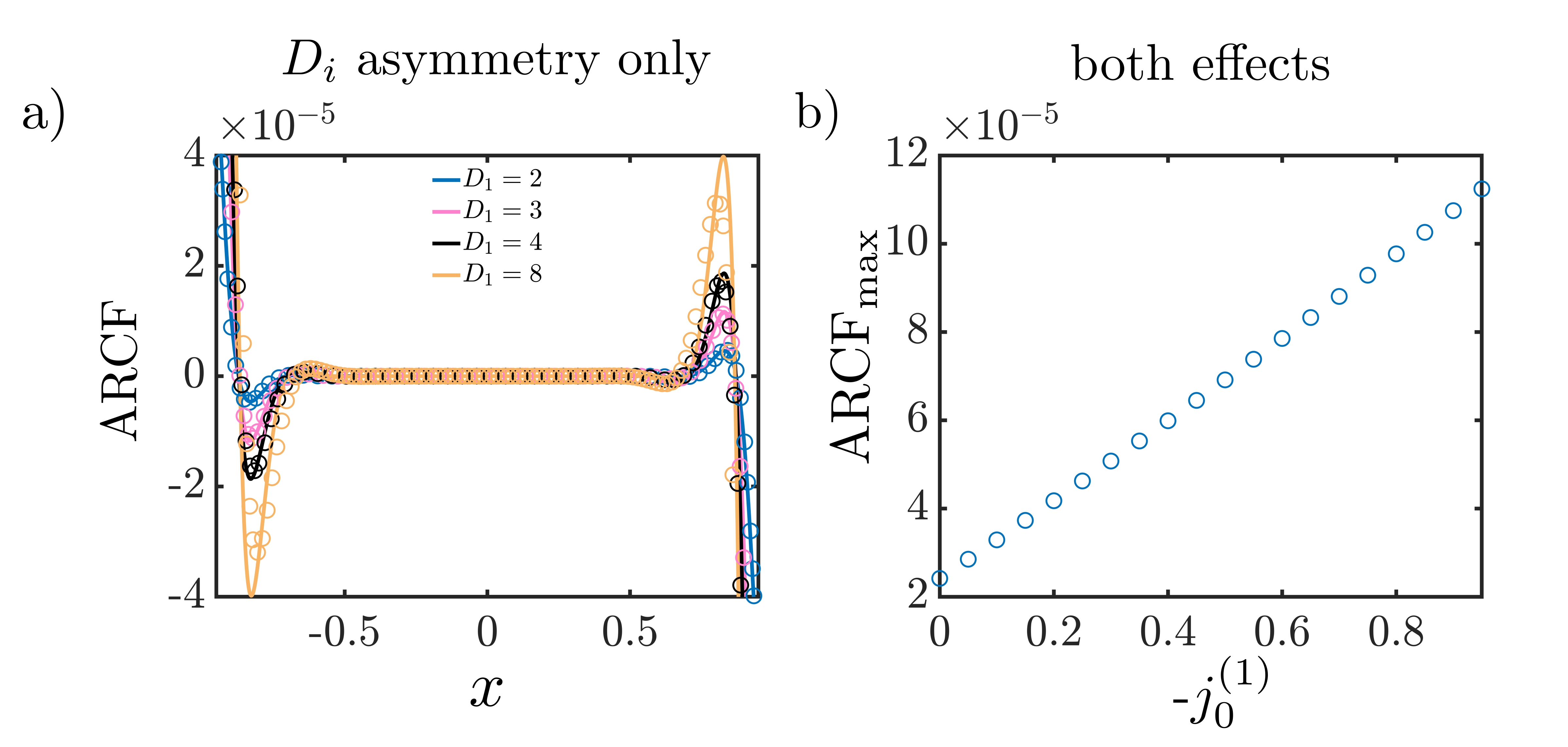}
    \caption{Presence of ARCFs.  (a) ARCF vs $x$ for diffusivity asymmetry for $D_1 = 2$ (blue), $D_1 = 3$ (pink), $D_1 = 4$ (black), and $D_1 = 8$ (orange). $D_2 = 1$ in all cases. (b) ARCF$_{\textrm{max}}$ vs $-j_0^{(1)}$ with both diffusivity asymmetry and surface reactions (only cation reactive). $D_1 = 1$ and $D_2 = 4$. In all cases, analytical results are solid lines and numerical simulations are orbs.  $\kappa = 100$, $\omega = 1000$, $z_1 = 1$, and $z_2 = -1$.  }
    \label{Fig: Fig8}
\end{figure*}

\subsection{Asymmetric Rectified Concentration Fields (ARCFs)}
\label{sec: arcf}
We investigate the formation of ARCFs in a binary electrolyte.  We employ Eq. (\ref{eq: secondordersaltfinal}) to estimate the ARCF based on first-order results from (i) numerical calculations described in section \ref{subsec: numericsprocedure} and (ii) analytical calculations described in section \ref{sec: analytics} and the Appendix. 
\par{} We note that ARCFs form when $\hat{\rho}_e^{(1)} \neq 0$ beyond the EDLs. To this end, we find that surface reactions alone are unable to produce ARCFs in the limit $\sqrt{I^{(0)}}\kappa \gg \sqrt{\omega}$, which is valid in experiments \cite{woehl2015bifurcation, bukosky_extreme_2019, wang2022long, rath2021ph}. In this limit, for symmetric diffusivities, Eq. (\ref{eq: analyticsrho1eqn1}) reduces to $\hat{\rho}_e^{(1)} \approx - \left( j_0^{(1)} + I^{(0)} \right) \frac{\sinh \left( \sqrt{I^{(0)}} \kappa x \right)} {\sinh \left( \sqrt{I^{(0)}} \kappa  \right) }$, which suggests that the charge is only accumulated in the EDLs and ARCFs do not form. We note that Bazant et al. \cite{bazant2004diffuse} demonstrated that a region outside the EDLs could accumulate charge for large potentials even for symmetric diffusivities, but this effect cannot be captured in our analysis and could be a potential avenue for future research. 

\par{} Now, we shift our focus to the case of asymmetric diffusivities only. We show that $\hat{\rho}_e^{(1)} \neq 0$ outside EDLs for $D_1 \neq D_2$; see Fig. \ref{Fig: Fig8}(a). As expected, the magnitude of $\hat{\rho}_e^{(1)}$ increases with an increase in $D_1$. This charge imbalance occurs because the equations for the charge and salt are coupled for asymmetric diffusivities \cite{hashemi_asymmetric_2020, henrique2022charging}; also see the Appendix. The induced ARCFs due to these charge imbalances are shown in Fig. \ref{Fig: Fig8}(a). The ARCF is also a long-range steady field, indicating that it could be important for relatively large distances away from the electrodes. Interestingly, the spatial dependency of the ARCF remains identical even if the values of $D_1$ and $D_2$ are interchanged (results not shown), in contrast to AREFs \cite{hashemi_oscillating_2018, hashemi_asymmetric_2020, balu_thin_2021}. We note that the magnitude of the ARCF is smaller compared to the AREF; see Fig. \ref{Fig: Fig2}. While this might suggest that the role of the ARCF is minor, we note that the shape of the ARCF is different than the AREF and could thus influence the regions where the AREF is smaller.  We also anticipate that ARCFs will become stronger in the nonlinear applied potential limit \cite{bazant2004diffuse}.  Furthermore, the relative importance of AREFs and ARCFs will depend on the interactions between the ionic species and particles \cite{anderson1989colloid}. This is particularly important for the phenomena of electrodiffusiophoresis, where both electrophoresis and diffusiophoresis are present \cite{wang2022long, wang2022visualization}. We emphasize that while surface reactions cannot induce ARCFs on their own, they can enhance ARCFs caused by diffusivity asymmetry; see Fig. \ref{Fig: Fig8}(b), which shows the linear dependency of $\textrm{ARCF}_{\textrm{max}}$ on $j_0^{(1)}$.

\subsection{Validity of the proposed framework}
Since our work assumes $\phi_D \ll 1 $, the results are most accurate up to an applied potential of $\pm 25$ mV. However, as shown by Balu and Khair \cite{balu_thin_2021}, the peak value of AREFs is relatively linear up to $\phi_D = 10$. This means that the results shown here may be extra-polated until applied potentials of $\pm 250$ mV with a low to moderate loss in accuracy. With even larger voltages, this framework is only suitable for a qualitative analysis. The surface reactive flux boundary conditions used in this work break down when reaction rate is no longer linear with potential; the potential value at which they break down will depend on the transfer coefficients and Stern layer thickness \cite{biesheuvel2011diffuse}, and we thus refrain from making a quantitative estimate of this limit.

\subsection{Limitations of the proposed framework}
\label{sec: limitation}
The proposed framework demonstrates that electrochemical reactions can produce AREFs and that ARCFs are also present in the system. Further, our work elucidates that imbalances in ionic strength and charge density outside the EDLs produce AREFs and ARCFs, respectively. Nonetheless, our work has two primary limitations. 
\par{} The first limitation of this work is that it assumes a small applied potential. In experiments, the voltage applied is significantly larger \cite{woehl2015bifurcation, bukosky_extreme_2019, wang2022long, wang2022visualization}, and thus the nonlinear PNP equations would need to be solved. In this scenario, typically, the thin EDL limit is invoked and a singular perturbation expansion is performed \cite{bazant2004diffuse, balu_role_2018, balu_thin_2021, balu2022electrochemical, jarvey2022ion}. The analysis for a binary electrolyte with asymmetric diffusivities and no electrochemical reactions has been investigated by Balu and Khair \cite{balu_thin_2021}. Based on the trends observed in Fig, \ref{Fig: Fig2}(b), we anticipate that the AREFs due to chemical reactions will appear at leading- and first-order expansion terms in the singular perturbation expansion, unlike asymmetric diffusivities where they are observed at the second-order \cite{balu_thin_2021}, though a complete analysis is required to be certain. For a system with electrochemical reactions and symmetric diffusivities, the framework proposed in our prior work \cite{jarvey2022ion} can be extended.  
\par{} The second limitation of our work is that it ignores the effect of equilibrium cell potential. Clearly, the surface reactions become significant beyond a certain cell voltage. To capture such an effect, Frumkin-Butler-Volmer kinetics could be invoked \cite{bazant2005current,biesheuvel2011diffuse}.  In this approach, forward and reverse reactions are written in terms of overpotential, defined as the difference between the applied potential and the equilibrium cell potential. Additionally, a Stern layer needs to be accounted for in the system to write the Frumkin-Butler-Volmer kinetic equation. There are two primary differences from the approach described in this paper. First, the leading order solution requires the formation of equilibrium double layers. Second, the boundary conditions for potential will now be written as a linear potential drop across each Stern layer.  We note that our prior work \cite{jarvey2022ion} provides a methodology to capture these effects, while also simultaneously capturing the effect of EDLs and the region outside EDLs, albeit in the limit of symmetric diffusivities. To capture these effects for asymmetric diffusivities, the frameworks laid out by Balu and Khair \cite{balu_thin_2021} and Jarvey et al. \cite{jarvey2022ion} would need to be combined, but the analysis is likely to be limited to a binary electrolyte. For multicomponent electrolytes, numerical calculations would be required.

\par{} We emphasize that while the opportunities outlined above will improve the accuracy of the current results, qualitatively, the small potential calculations are able to capture the essential physics of the system.  
\section{Conclusion and Outlook}
\label{sec: outlook}
In this article, we performed a regular perturbation expansion in the small-applied-potential limit on the Poisson-Nernst-Planck equations for multicomponent electrolytes for arbitrary diffusivities and valences, while also including the effect of electrochemical surface reactions. Our results highlight  that surface electrochemical reactions are an additional mechanism for AREF formation. We show that an imbalance in ionic strength is a prerequisite for a nonzero AREF. Further, we find that ARCFs may also be present in electrochemical cells and could induce a diffusiophoretic force on the particles. We show analytical expressions for AREFs and ARCFs are possible by further invoking the thin-double-layer limit. AREFs caused by surface reactions are less sensitive to parameters as compared to AREFs from diffusivity contrast. Lastly, we find that ARCFs appear primarily due to diffusivity contrast, though electrochemical reactions can enhance them.

\par{} Our contributions are directly applicable for the directed assembly of colloids using electric fields, including ellipsoids \cite{mittal2009electric}, colloidal dumbbells \cite{demirors2010directed}, colloidal dimers \cite{ma2012two}, dicolloids \cite{panczyk2013two}, Janus particles \cite{zhang2012directed}, patchy anisotropic microparticles, \cite{shields2013field} and chiral clusters \cite{ma2015electric}, among others. This assembly method holds great promise in creating materials with superior optical and electrical properties \cite{shields2013field, ma2012two, ma2015electric, kruglova_wonders_2013}, such as photonic crystals, microactuators, and colloidal robots. The work presented here can help estimate the assembly of colloids in an electrochemical cell, where reactions provide an additional knob to tune colloidal assembly \cite{silvera_batista_controlled_2017, wang2022long, wang2022visualization, rath2021ph}. 
\par{} We highlight that the implications of our findings extend beyond the realm of colloids. Specifically, multicomponent electrolytes and surface reactions are often used in batteries \cite{li201830,zhang2006review}, Faradaic desalination \cite{biesheuvel_electrochemistry_2012,sayed_faradic_2021}, carbon dioxide reduction \cite{wu2017co2}, hybrid capacitors \cite{simon2020perspectives}, and reversible metal electrodeposition windows \cite{tao2021reversible}. While multicomponent electrolytes and surface electrochemical reactions are prevalent in experimental literature, theoretical understanding of ion transport in these systems still remains elusive. In our previous work \cite{jarvey2022ion}, we analyzed the coupled effects of electrical double layers and surface electrochemical reactions on ion transport in multicomponent electrolytes for a DC potential. This work furthers the literature on AC fields, which have important implications in electrochemical impedance spectroscopy \cite{balu2022electrochemical} and transport in porous materials \cite{de1963porous, de1964porous, henrique2022impact, henrique2022charging}. The results outlined here are also a potential avenue for improving upon the modified Donnan potential approach \cite{biesheuvel2011diffuse, biesheuvel_electrochemistry_2012} for modeling Faradaic capacitive deionization \cite{he2018theory}.

\section{Acknowledgements}
\label{sec: acknowledgements}
A.G. thanks the National Science Foundation (CBET - 2238412) CAREER award for financial support. Acknowledgement is made to the donors of the American Chemical Society Petroleum Research Fund for partial support of this research. N. J. thanks the ARCS Foundation Scholarship and GAANN fellowship in Soft Materials  for financial support. F. H. acknowledges the Ryland Family Graduate Fellowship for financial assistance. 

\section*{Appendix: Adapting results from Hashemi et al. \cite{hashemi2020perturbation}}
\label{sec: appendix1}
At first order in $\phi_D$, we directly apply the analytical results of Hashemi et al. \cite{hashemi2020perturbation}. Note that their derivation is only valid for binary electrolytes, and as such is used exclusively to compare with results for binary electrolytes. To match the formulation used in their work, we define
\begin{subequations}
\begin{eqnarray}
\kappa_h = \kappa \sqrt{z_2^2 z_1 - z_1^2 z_2}, \\
\omega_h = \frac{\omega}{\sqrt{2} \kappa^2}, \\
\beta = \frac{D_1 - D_2}{D_1 + D_2}, \\
\gamma = \frac{z_1 + z_2}{z_1 - z_2},
\end{eqnarray}
\end{subequations}
where ion 1 is the cation and ion 2 is the anion. We then write the additional parameters in their analytical solutions such that
\begin{subequations}
\begin{eqnarray}
\Delta = 1 - 4 \beta \omega_h \left(i \gamma + \beta \omega_h \right), \\
s = 2 i \beta \omega_h + \sqrt{\Delta}, \\
z = 0.5 \left(z_1 - z_2 \right), \\
\lambda_+ = \sqrt{\frac{1+ 2i \omega_h + \sqrt{\Delta}}{2}}, \\
\lambda_- = \sqrt{\frac{1+ 2i \omega_h - \sqrt{\Delta}}{2}}.
\end{eqnarray}
\end{subequations}
Continuing to build to the forms of the first-order solutions the authors arrive at, we write
\begin{subequations}
\begin{eqnarray}
\Gamma = s^2 + 2\gamma + 1 - \frac{1}{2 \kappa_h}\left( \frac{(\gamma+1) (s-1)^2 (\lambda_- \kappa_h -\tanh{\lambda_- \kappa_h})}{\lambda_-^3}  \right. - \\ \left. \frac{(\gamma-1) (s+1)^2 (\lambda_+ \kappa_h -\tanh{\lambda_+ \kappa_h})}{\lambda_+^3} \right), \nonumber \\
A = \frac{s - 1}{\lambda_- \kappa_h \cosh{\left( \lambda_- \kappa_h\right)} \Gamma}, \\
\mathcal{B} =\frac{s + 1}{\lambda_+ \kappa_h \cosh{\left( \lambda_+ \kappa_h\right)} \Gamma} , \\
C = \kappa_h^{-1} \left(-1 + \frac{A (1 + \gamma) (s - 1) \sinh{\lambda_- \kappa_h}}{2 \lambda_-^2} + \right. \\ \left. \frac{\mathcal{B} (1 - \gamma) (s + 1) \sinh{\lambda_+ \kappa_h}}{2 \lambda_+^2} \right). \nonumber
\end{eqnarray}
\end{subequations}
Finally, we write the first order variables
\begin{subequations}
\begin{eqnarray}
\hat{c}_1^{(1)} = -\left( A (-\gamma + s) \sinh{\left( \lambda_- x \kappa_h \right) } + \mathcal{B} (1 + \gamma) \sinh{\left( \lambda_+ x \kappa_h \right)}  \right), \\
\hat{c}_2^{(1)} = -\left( A (1 + \gamma) \sinh{\left( \lambda_- x \kappa_h \right) } - \mathcal{B} (-\gamma + s) \sinh{\left( \lambda_+ x \kappa_h \right)}  \right), \\
\hat{\phi}^{(1)} = - z^{-1} \left(C x \kappa_h - \frac{A (1+\gamma) (s-1) \sinh{\left(\lambda_- x \kappa  \right)}}{2 \lambda_-^2} \right. - \\ \left. \frac{\mathcal{B} (1-\gamma) (s+1) \sinh{\left(\lambda_+ x \kappa  \right)}}{2 \lambda_+^2} \right), \nonumber \\
\label{eq: hashemianalyticsfinal}
\hat{E}^{(1)} = - z^{-1} \left( C \kappa_h - \lambda_- \kappa_h \frac{A (1+\gamma) (s-1) \cosh{\left(\lambda_- x \kappa  \right)}}{2 \lambda_-^2} - \right. \\ \left. \lambda_+ \kappa_h \frac{\mathcal{B} (1-\gamma) (s+1) \sinh{\left(\lambda_+ x \kappa  \right)}}{2 \lambda_+^2} \right). \nonumber
\end{eqnarray}
\end{subequations}
\par{} To determine the AREF and ARCF from these first-order results, we employ Eqs. (\ref{eq: analyticsAREFfinal}) and (\ref{eq: secondordersaltfinal}). Note that the remainder of the analytics in section \ref{sec: analytics} do not hold for asymmetric diffusivities, but Eqs. (\ref{eq: analyticsAREFfinal}) and (\ref{eq: secondordersaltfinal}) are valid for asymmetric diffusivities.



\renewcommand\refname{References}

\bibliography{apssamp}

\begin{thebibliography}{66}%
\makeatletter
\providecommand \@ifxundefined [1]{%
 \@ifx{#1\undefined}
}%
\providecommand \@ifnum [1]{%
 \ifnum #1\expandafter \@firstoftwo
 \else \expandafter \@secondoftwo
 \fi
}%
\providecommand \@ifx [1]{%
 \ifx #1\expandafter \@firstoftwo
 \else \expandafter \@secondoftwo
 \fi
}%
\providecommand \natexlab [1]{#1}%
\providecommand \enquote  [1]{``#1''}%
\providecommand \bibnamefont  [1]{#1}%
\providecommand \bibfnamefont [1]{#1}%
\providecommand \citenamefont [1]{#1}%
\providecommand \href@noop [0]{\@secondoftwo}%
\providecommand \href [0]{\begingroup \@sanitize@url \@href}%
\providecommand \@href[1]{\@@startlink{#1}\@@href}%
\providecommand \@@href[1]{\endgroup#1\@@endlink}%
\providecommand \@sanitize@url [0]{\catcode `\\12\catcode `\$12\catcode
  `\&12\catcode `\#12\catcode `\^12\catcode `\_12\catcode `\%12\relax}%
\providecommand \@@startlink[1]{}%
\providecommand \@@endlink[0]{}%
\providecommand \url  [0]{\begingroup\@sanitize@url \@url }%
\providecommand \@url [1]{\endgroup\@href {#1}{\urlprefix }}%
\providecommand \urlprefix  [0]{URL }%
\providecommand \Eprint [0]{\href }%
\providecommand \doibase [0]{https://doi.org/}%
\providecommand \selectlanguage [0]{\@gobble}%
\providecommand \bibinfo  [0]{\@secondoftwo}%
\providecommand \bibfield  [0]{\@secondoftwo}%
\providecommand \translation [1]{[#1]}%
\providecommand \BibitemOpen [0]{}%
\providecommand \bibitemStop [0]{}%
\providecommand \bibitemNoStop [0]{.\EOS\space}%
\providecommand \EOS [0]{\spacefactor3000\relax}%
\providecommand \BibitemShut  [1]{\csname bibitem#1\endcsname}%
\let\auto@bib@innerbib\@empty
\bibitem [{\citenamefont {Sides}(2001)}]{sides2001electrohydrodynamic}%
  \BibitemOpen
  \bibfield  {author} {\bibinfo {author} {\bibfnamefont {P.~J.}\ \bibnamefont
  {Sides}},\ }\bibfield  {title} {\bibinfo {title} {Electrohydrodynamic
  particle aggregation on an electrode driven by an alternating electric field
  normal to it},\ }\href@noop {} {\bibfield  {journal} {\bibinfo  {journal}
  {Langmuir}\ }\textbf {\bibinfo {volume} {17}},\ \bibinfo {pages} {5791}
  (\bibinfo {year} {2001})}\BibitemShut {NoStop}%
\bibitem [{\citenamefont {Ristenpart}\ \emph {et~al.}(2007)\citenamefont
  {Ristenpart}, \citenamefont {Aksay},\ and\ \citenamefont
  {Saville}}]{ristenpart2007electrohydrodynamic}%
  \BibitemOpen
  \bibfield  {author} {\bibinfo {author} {\bibfnamefont {W.}~\bibnamefont
  {Ristenpart}}, \bibinfo {author} {\bibfnamefont {I.~A.}\ \bibnamefont
  {Aksay}},\ and\ \bibinfo {author} {\bibfnamefont {D.}~\bibnamefont
  {Saville}},\ }\bibfield  {title} {\bibinfo {title} {Electrohydrodynamic flow
  around a colloidal particle near an electrode with an oscillating
  potential},\ }\href@noop {} {\bibfield  {journal} {\bibinfo  {journal} {J.
  Fluid Mech.}\ }\textbf {\bibinfo {volume} {575}},\ \bibinfo {pages} {83}
  (\bibinfo {year} {2007})}\BibitemShut {NoStop}%
\bibitem [{\citenamefont {Wirth}\ \emph {et~al.}(2011)\citenamefont {Wirth},
  \citenamefont {Rock}, \citenamefont {Sides},\ and\ \citenamefont
  {Prieve}}]{wirth2011single}%
  \BibitemOpen
  \bibfield  {author} {\bibinfo {author} {\bibfnamefont {C.~L.}\ \bibnamefont
  {Wirth}}, \bibinfo {author} {\bibfnamefont {R.~M.}\ \bibnamefont {Rock}},
  \bibinfo {author} {\bibfnamefont {P.~J.}\ \bibnamefont {Sides}},\ and\
  \bibinfo {author} {\bibfnamefont {D.~C.}\ \bibnamefont {Prieve}},\ }\bibfield
   {title} {\bibinfo {title} {Single and pairwise motion of particles near an
  ideally polarizable electrode},\ }\href@noop {} {\bibfield  {journal}
  {\bibinfo  {journal} {Langmuir}\ }\textbf {\bibinfo {volume} {27}},\ \bibinfo
  {pages} {9781} (\bibinfo {year} {2011})}\BibitemShut {NoStop}%
\bibitem [{\citenamefont {Wirth}\ \emph {et~al.}(2013)\citenamefont {Wirth},
  \citenamefont {Sides},\ and\ \citenamefont {Prieve}}]{wirth2013electrolyte}%
  \BibitemOpen
  \bibfield  {author} {\bibinfo {author} {\bibfnamefont {C.~L.}\ \bibnamefont
  {Wirth}}, \bibinfo {author} {\bibfnamefont {P.~J.}\ \bibnamefont {Sides}},\
  and\ \bibinfo {author} {\bibfnamefont {D.~C.}\ \bibnamefont {Prieve}},\
  }\bibfield  {title} {\bibinfo {title} {Electrolyte dependence of particle
  motion near an electrode during ac polarization},\ }\href@noop {} {\bibfield
  {journal} {\bibinfo  {journal} {Phys. Rev. E}\ }\textbf {\bibinfo {volume}
  {87}},\ \bibinfo {pages} {032302} (\bibinfo {year} {2013})}\BibitemShut
  {NoStop}%
\bibitem [{\citenamefont {Hoggard}\ \emph {et~al.}(2008)\citenamefont
  {Hoggard}, \citenamefont {Sides},\ and\ \citenamefont
  {Prieve}}]{hoggard2008electrolyte}%
  \BibitemOpen
  \bibfield  {author} {\bibinfo {author} {\bibfnamefont {J.~D.}\ \bibnamefont
  {Hoggard}}, \bibinfo {author} {\bibfnamefont {P.~J.}\ \bibnamefont {Sides}},\
  and\ \bibinfo {author} {\bibfnamefont {D.~C.}\ \bibnamefont {Prieve}},\
  }\bibfield  {title} {\bibinfo {title} {Electrolyte-dependent multiparticle
  motion near electrodes in oscillating electric fields},\ }\href@noop {}
  {\bibfield  {journal} {\bibinfo  {journal} {Langmuir}\ }\textbf {\bibinfo
  {volume} {24}},\ \bibinfo {pages} {2977} (\bibinfo {year}
  {2008})}\BibitemShut {NoStop}%
\bibitem [{\citenamefont {Prieve}\ \emph {et~al.}(2010)\citenamefont {Prieve},
  \citenamefont {Sides},\ and\ \citenamefont {Wirth}}]{prieve_2-d_2010}%
  \BibitemOpen
  \bibfield  {author} {\bibinfo {author} {\bibfnamefont {D.~C.}\ \bibnamefont
  {Prieve}}, \bibinfo {author} {\bibfnamefont {P.~J.}\ \bibnamefont {Sides}},\
  and\ \bibinfo {author} {\bibfnamefont {C.~L.}\ \bibnamefont {Wirth}},\
  }\bibfield  {title} {\bibinfo {title} {2-{D} assembly of colloidal particles
  on a planar electrode},\ }\href {https://doi.org/10.1016/j.cocis.2010.01.005}
  {\bibfield  {journal} {\bibinfo  {journal} {Curr. Opin. Colloid Interface
  Sci.}\ }\textbf {\bibinfo {volume} {15}},\ \bibinfo {pages} {160} (\bibinfo
  {year} {2010})}\BibitemShut {NoStop}%
\bibitem [{\citenamefont {Woehl}\ \emph {et~al.}(2014)\citenamefont {Woehl},
  \citenamefont {Heatley}, \citenamefont {Dutcher}, \citenamefont {Talken},\
  and\ \citenamefont {Ristenpart}}]{woehl_electrolyte-dependent_2014}%
  \BibitemOpen
  \bibfield  {author} {\bibinfo {author} {\bibfnamefont {T.~J.}\ \bibnamefont
  {Woehl}}, \bibinfo {author} {\bibfnamefont {K.~L.}\ \bibnamefont {Heatley}},
  \bibinfo {author} {\bibfnamefont {C.~S.}\ \bibnamefont {Dutcher}}, \bibinfo
  {author} {\bibfnamefont {N.~H.}\ \bibnamefont {Talken}},\ and\ \bibinfo
  {author} {\bibfnamefont {W.~D.}\ \bibnamefont {Ristenpart}},\ }\bibfield
  {title} {\bibinfo {title} {Electrolyte-{Dependent} {Aggregation} of
  {Colloidal} {Particles} near {Electrodes} in {Oscillatory} {Electric}
  {Fields}},\ }\href {https://doi.org/10.1021/la4048243} {\bibfield  {journal}
  {\bibinfo  {journal} {Langmuir}\ }\textbf {\bibinfo {volume} {30}},\ \bibinfo
  {pages} {4887} (\bibinfo {year} {2014})},\ \bibinfo {note} {publisher:
  American Chemical Society}\BibitemShut {NoStop}%
\bibitem [{\citenamefont {Saini}\ \emph {et~al.}(2016)\citenamefont {Saini},
  \citenamefont {Bukosky},\ and\ \citenamefont
  {Ristenpart}}]{saini2016influence}%
  \BibitemOpen
  \bibfield  {author} {\bibinfo {author} {\bibfnamefont {S.}~\bibnamefont
  {Saini}}, \bibinfo {author} {\bibfnamefont {S.~C.}\ \bibnamefont {Bukosky}},\
  and\ \bibinfo {author} {\bibfnamefont {W.~D.}\ \bibnamefont {Ristenpart}},\
  }\bibfield  {title} {\bibinfo {title} {Influence of electrolyte concentration
  on the aggregation of colloidal particles near electrodes in oscillatory
  fields},\ }\href@noop {} {\bibfield  {journal} {\bibinfo  {journal}
  {Langmuir}\ }\textbf {\bibinfo {volume} {32}},\ \bibinfo {pages} {4210}
  (\bibinfo {year} {2016})}\BibitemShut {NoStop}%
\bibitem [{\citenamefont {Ristenpart}\ \emph {et~al.}(2004)\citenamefont
  {Ristenpart}, \citenamefont {Aksay},\ and\ \citenamefont
  {Saville}}]{ristenpart2004assembly}%
  \BibitemOpen
  \bibfield  {author} {\bibinfo {author} {\bibfnamefont {W.}~\bibnamefont
  {Ristenpart}}, \bibinfo {author} {\bibfnamefont {I.~A.}\ \bibnamefont
  {Aksay}},\ and\ \bibinfo {author} {\bibfnamefont {D.}~\bibnamefont
  {Saville}},\ }\bibfield  {title} {\bibinfo {title} {Assembly of colloidal
  aggregates by electrohydrodynamic flow: Kinetic experiments and scaling
  analysis},\ }\href@noop {} {\bibfield  {journal} {\bibinfo  {journal} {Phys.
  Rev. E}\ }\textbf {\bibinfo {volume} {69}},\ \bibinfo {pages} {021405}
  (\bibinfo {year} {2004})}\BibitemShut {NoStop}%
\bibitem [{\citenamefont {Woehl}\ \emph {et~al.}(2015)\citenamefont {Woehl},
  \citenamefont {Chen}, \citenamefont {Heatley}, \citenamefont {Talken},
  \citenamefont {Bukosky}, \citenamefont {Dutcher},\ and\ \citenamefont
  {Ristenpart}}]{woehl2015bifurcation}%
  \BibitemOpen
  \bibfield  {author} {\bibinfo {author} {\bibfnamefont {T.}~\bibnamefont
  {Woehl}}, \bibinfo {author} {\bibfnamefont {B.}~\bibnamefont {Chen}},
  \bibinfo {author} {\bibfnamefont {K.}~\bibnamefont {Heatley}}, \bibinfo
  {author} {\bibfnamefont {N.}~\bibnamefont {Talken}}, \bibinfo {author}
  {\bibfnamefont {S.}~\bibnamefont {Bukosky}}, \bibinfo {author} {\bibfnamefont
  {C.}~\bibnamefont {Dutcher}},\ and\ \bibinfo {author} {\bibfnamefont
  {W.}~\bibnamefont {Ristenpart}},\ }\bibfield  {title} {\bibinfo {title}
  {Bifurcation in the steady-state height of colloidal particles near an
  electrode in oscillatory electric fields: Evidence for a tertiary potential
  minimum},\ }\href@noop {} {\bibfield  {journal} {\bibinfo  {journal} {Phys.
  Rev. X}\ }\textbf {\bibinfo {volume} {5}},\ \bibinfo {pages} {011023}
  (\bibinfo {year} {2015})}\BibitemShut {NoStop}%
\bibitem [{\citenamefont {Bukosky}\ and\ \citenamefont
  {Ristenpart}(2015)}]{bukosky2015simultaneous}%
  \BibitemOpen
  \bibfield  {author} {\bibinfo {author} {\bibfnamefont {S.~C.}\ \bibnamefont
  {Bukosky}}\ and\ \bibinfo {author} {\bibfnamefont {W.~D.}\ \bibnamefont
  {Ristenpart}},\ }\bibfield  {title} {\bibinfo {title} {Simultaneous
  aggregation and height bifurcation of colloidal particles near electrodes in
  oscillatory electric fields},\ }\href@noop {} {\bibfield  {journal} {\bibinfo
   {journal} {Langmuir}\ }\textbf {\bibinfo {volume} {31}},\ \bibinfo {pages}
  {9742} (\bibinfo {year} {2015})}\BibitemShut {NoStop}%
\bibitem [{\citenamefont {Hashemi}\ \emph {et~al.}(2018)\citenamefont
  {Hashemi}, \citenamefont {Bukosky}, \citenamefont {Rader}, \citenamefont
  {Ristenpart},\ and\ \citenamefont {Miller}}]{hashemi_oscillating_2018}%
  \BibitemOpen
  \bibfield  {author} {\bibinfo {author} {\bibfnamefont {A.}~\bibnamefont
  {Hashemi}}, \bibinfo {author} {\bibfnamefont {S.~C.}\ \bibnamefont
  {Bukosky}}, \bibinfo {author} {\bibfnamefont {S.~P.}\ \bibnamefont {Rader}},
  \bibinfo {author} {\bibfnamefont {W.~D.}\ \bibnamefont {Ristenpart}},\ and\
  \bibinfo {author} {\bibfnamefont {G.~H.}\ \bibnamefont {Miller}},\ }\bibfield
   {title} {\bibinfo {title} {Oscillating {Electric} {Fields} in {Liquids}
  {Create} a {Long}-{Range} {Steady} {Field}},\ }\href
  {https://doi.org/10.1103/PhysRevLett.121.185504} {\bibfield  {journal}
  {\bibinfo  {journal} {Phys. Rev. Lett.}\ }\textbf {\bibinfo {volume} {121}},\
  \bibinfo {pages} {185504} (\bibinfo {year} {2018})},\ \bibinfo {note}
  {publisher: American Physical Society}\BibitemShut {NoStop}%
\bibitem [{\citenamefont {Hashemi}\ \emph {et~al.}(2019)\citenamefont
  {Hashemi}, \citenamefont {Miller},\ and\ \citenamefont
  {Ristenpart}}]{hashemi_asymmetric_2019}%
  \BibitemOpen
  \bibfield  {author} {\bibinfo {author} {\bibfnamefont {A.}~\bibnamefont
  {Hashemi}}, \bibinfo {author} {\bibfnamefont {G.~H.}\ \bibnamefont
  {Miller}},\ and\ \bibinfo {author} {\bibfnamefont {W.~D.}\ \bibnamefont
  {Ristenpart}},\ }\bibfield  {title} {\bibinfo {title} {Asymmetric rectified
  electric fields between parallel electrodes: {Numerical} and scaling
  analyses},\ }\href {https://doi.org/10.1103/PhysRevE.99.062603} {\bibfield
  {journal} {\bibinfo  {journal} {Phys. Rev. E}\ }\textbf {\bibinfo {volume}
  {99}},\ \bibinfo {pages} {062603} (\bibinfo {year} {2019})},\ \bibinfo {note}
  {publisher: American Physical Society}\BibitemShut {NoStop}%
\bibitem [{\citenamefont {Bukosky}\ \emph {et~al.}(2019)\citenamefont
  {Bukosky}, \citenamefont {Hashemi}, \citenamefont {Rader}, \citenamefont
  {Mora}, \citenamefont {Miller},\ and\ \citenamefont
  {Ristenpart}}]{bukosky_extreme_2019}%
  \BibitemOpen
  \bibfield  {author} {\bibinfo {author} {\bibfnamefont {S.~C.}\ \bibnamefont
  {Bukosky}}, \bibinfo {author} {\bibfnamefont {A.}~\bibnamefont {Hashemi}},
  \bibinfo {author} {\bibfnamefont {S.~P.}\ \bibnamefont {Rader}}, \bibinfo
  {author} {\bibfnamefont {J.}~\bibnamefont {Mora}}, \bibinfo {author}
  {\bibfnamefont {G.~H.}\ \bibnamefont {Miller}},\ and\ \bibinfo {author}
  {\bibfnamefont {W.~D.}\ \bibnamefont {Ristenpart}},\ }\bibfield  {title}
  {\bibinfo {title} {Extreme {Levitation} of {Colloidal} {Particles} in
  {Response} to {Oscillatory} {Electric} {Fields}},\ }\href
  {https://doi.org/10.1021/acs.langmuir.9b00313} {\bibfield  {journal}
  {\bibinfo  {journal} {Langmuir}\ }\textbf {\bibinfo {volume} {35}},\ \bibinfo
  {pages} {6971} (\bibinfo {year} {2019})},\ \bibinfo {note} {publisher:
  American Chemical Society}\BibitemShut {NoStop}%
\bibitem [{\citenamefont {Hashemi}\ \emph
  {et~al.}(2020{\natexlab{a}})\citenamefont {Hashemi}, \citenamefont {Miller},\
  and\ \citenamefont {Ristenpart}}]{hashemi_asymmetric_2020}%
  \BibitemOpen
  \bibfield  {author} {\bibinfo {author} {\bibfnamefont {A.}~\bibnamefont
  {Hashemi}}, \bibinfo {author} {\bibfnamefont {G.~H.}\ \bibnamefont
  {Miller}},\ and\ \bibinfo {author} {\bibfnamefont {W.~D.}\ \bibnamefont
  {Ristenpart}},\ }\bibfield  {title} {\bibinfo {title} {Asymmetric rectified
  electric fields generate flows that can dominate induced-charge
  electrokinetics},\ }\href {https://doi.org/10.1103/PhysRevFluids.5.013702}
  {\bibfield  {journal} {\bibinfo  {journal} {Phys. Rev. Fluids}\ }\textbf
  {\bibinfo {volume} {5}},\ \bibinfo {pages} {013702} (\bibinfo {year}
  {2020}{\natexlab{a}})},\ \bibinfo {note} {publisher: American Physical
  Society}\BibitemShut {NoStop}%
\bibitem [{\citenamefont {Mirzadeh}\ and\ \citenamefont
  {Gibou}(2014)}]{mirzadeh2014conservative}%
  \BibitemOpen
  \bibfield  {author} {\bibinfo {author} {\bibfnamefont {M.}~\bibnamefont
  {Mirzadeh}}\ and\ \bibinfo {author} {\bibfnamefont {F.}~\bibnamefont
  {Gibou}},\ }\bibfield  {title} {\bibinfo {title} {A conservative
  discretization of the poisson--nernst--planck equations on adaptive cartesian
  grids},\ }\href@noop {} {\bibfield  {journal} {\bibinfo  {journal} {J. Comp.
  Phys.}\ }\textbf {\bibinfo {volume} {274}},\ \bibinfo {pages} {633} (\bibinfo
  {year} {2014})}\BibitemShut {NoStop}%
\bibitem [{\citenamefont {Hashemi}\ \emph
  {et~al.}(2020{\natexlab{b}})\citenamefont {Hashemi}, \citenamefont {Miller},
  \citenamefont {Bishop},\ and\ \citenamefont
  {Ristenpart}}]{hashemi2020perturbation}%
  \BibitemOpen
  \bibfield  {author} {\bibinfo {author} {\bibfnamefont {A.}~\bibnamefont
  {Hashemi}}, \bibinfo {author} {\bibfnamefont {G.~H.}\ \bibnamefont {Miller}},
  \bibinfo {author} {\bibfnamefont {K.~J.}\ \bibnamefont {Bishop}},\ and\
  \bibinfo {author} {\bibfnamefont {W.~D.}\ \bibnamefont {Ristenpart}},\
  }\bibfield  {title} {\bibinfo {title} {A perturbation solution to the full
  poisson--nernst--planck equations yields an asymmetric rectified electric
  field},\ }\href@noop {} {\bibfield  {journal} {\bibinfo  {journal} {Soft
  Matter}\ }\textbf {\bibinfo {volume} {16}},\ \bibinfo {pages} {7052}
  (\bibinfo {year} {2020}{\natexlab{b}})}\BibitemShut {NoStop}%
\bibitem [{\citenamefont {Balu}\ and\ \citenamefont
  {Khair}(2021)}]{balu_thin_2021}%
  \BibitemOpen
  \bibfield  {author} {\bibinfo {author} {\bibfnamefont {B.}~\bibnamefont
  {Balu}}\ and\ \bibinfo {author} {\bibfnamefont {A.~S.}\ \bibnamefont
  {Khair}},\ }\bibfield  {title} {\bibinfo {title} {A thin double layer
  analysis of asymmetric rectified electric fields ({AREFs})},\ }\href
  {https://doi.org/10.1007/s10665-021-10139-x} {\bibfield  {journal} {\bibinfo
  {journal} {J. Eng. Math}\ }\textbf {\bibinfo {volume} {129}},\ \bibinfo
  {pages} {4} (\bibinfo {year} {2021})}\BibitemShut {NoStop}%
\bibitem [{\citenamefont {Silvera~Batista}\ \emph {et~al.}(2017)\citenamefont
  {Silvera~Batista}, \citenamefont {Rezvantalab}, \citenamefont {Larson},\ and\
  \citenamefont {Solomon}}]{silvera_batista_controlled_2017}%
  \BibitemOpen
  \bibfield  {author} {\bibinfo {author} {\bibfnamefont {C.~A.}\ \bibnamefont
  {Silvera~Batista}}, \bibinfo {author} {\bibfnamefont {H.}~\bibnamefont
  {Rezvantalab}}, \bibinfo {author} {\bibfnamefont {R.~G.}\ \bibnamefont
  {Larson}},\ and\ \bibinfo {author} {\bibfnamefont {M.~J.}\ \bibnamefont
  {Solomon}},\ }\bibfield  {title} {\bibinfo {title} {Controlled {Levitation}
  of {Colloids} through {Direct} {Current} {Electric} {Fields}},\ }\href
  {https://doi.org/10.1021/acs.langmuir.7b00835} {\bibfield  {journal}
  {\bibinfo  {journal} {Langmuir}\ }\textbf {\bibinfo {volume} {33}},\ \bibinfo
  {pages} {10861} (\bibinfo {year} {2017})},\ \bibinfo {note} {publisher:
  American Chemical Society}\BibitemShut {NoStop}%
\bibitem [{\citenamefont {Wang}\ \emph
  {et~al.}(2022{\natexlab{a}})\citenamefont {Wang}, \citenamefont {Leville},
  \citenamefont {Behdani},\ and\ \citenamefont {Batista}}]{wang2022long}%
  \BibitemOpen
  \bibfield  {author} {\bibinfo {author} {\bibfnamefont {K.}~\bibnamefont
  {Wang}}, \bibinfo {author} {\bibfnamefont {S.}~\bibnamefont {Leville}},
  \bibinfo {author} {\bibfnamefont {B.}~\bibnamefont {Behdani}},\ and\ \bibinfo
  {author} {\bibfnamefont {C.~A.~S.}\ \bibnamefont {Batista}},\ }\bibfield
  {title} {\bibinfo {title} {Long-range transport and directed assembly of
  charged colloids under aperiodic electrodiffusiophoresis},\ }\href@noop {}
  {\bibfield  {journal} {\bibinfo  {journal} {Soft Matter}\ }\textbf {\bibinfo
  {volume} {18}},\ \bibinfo {pages} {5949} (\bibinfo {year}
  {2022}{\natexlab{a}})}\BibitemShut {NoStop}%
\bibitem [{\citenamefont {Wang}\ \emph
  {et~al.}(2022{\natexlab{b}})\citenamefont {Wang}, \citenamefont {Behdani},\
  and\ \citenamefont {Silvera~Batista}}]{wang2022visualization}%
  \BibitemOpen
  \bibfield  {author} {\bibinfo {author} {\bibfnamefont {K.}~\bibnamefont
  {Wang}}, \bibinfo {author} {\bibfnamefont {B.}~\bibnamefont {Behdani}},\ and\
  \bibinfo {author} {\bibfnamefont {C.~A.}\ \bibnamefont {Silvera~Batista}},\
  }\bibfield  {title} {\bibinfo {title} {Visualization of concentration
  gradients and colloidal dynamics under electrodiffusiophoresis},\ }\href@noop
  {} {\bibfield  {journal} {\bibinfo  {journal} {Langmuir}\ }\textbf {\bibinfo
  {volume} {38}},\ \bibinfo {pages} {5663} (\bibinfo {year}
  {2022}{\natexlab{b}})}\BibitemShut {NoStop}%
\bibitem [{\citenamefont {Rath}\ \emph {et~al.}(2021)\citenamefont {Rath},
  \citenamefont {Weaver}, \citenamefont {Wang},\ and\ \citenamefont
  {Woehl}}]{rath2021ph}%
  \BibitemOpen
  \bibfield  {author} {\bibinfo {author} {\bibfnamefont {M.}~\bibnamefont
  {Rath}}, \bibinfo {author} {\bibfnamefont {J.}~\bibnamefont {Weaver}},
  \bibinfo {author} {\bibfnamefont {M.}~\bibnamefont {Wang}},\ and\ \bibinfo
  {author} {\bibfnamefont {T.}~\bibnamefont {Woehl}},\ }\bibfield  {title}
  {\bibinfo {title} {ph-mediated aggregation-to-separation transition for
  colloids near electrodes in oscillatory electric fields},\ }\href@noop {}
  {\bibfield  {journal} {\bibinfo  {journal} {Langmuir}\ }\textbf {\bibinfo
  {volume} {37}},\ \bibinfo {pages} {9346} (\bibinfo {year}
  {2021})}\BibitemShut {NoStop}%
\bibitem [{\citenamefont {Prieve}\ \emph {et~al.}(1984)\citenamefont {Prieve},
  \citenamefont {Anderson}, \citenamefont {Ebel},\ and\ \citenamefont
  {Lowell}}]{prieve1984motion}%
  \BibitemOpen
  \bibfield  {author} {\bibinfo {author} {\bibfnamefont {D.}~\bibnamefont
  {Prieve}}, \bibinfo {author} {\bibfnamefont {J.}~\bibnamefont {Anderson}},
  \bibinfo {author} {\bibfnamefont {J.}~\bibnamefont {Ebel}},\ and\ \bibinfo
  {author} {\bibfnamefont {M.}~\bibnamefont {Lowell}},\ }\bibfield  {title}
  {\bibinfo {title} {Motion of a particle generated by chemical gradients. part
  2. electrolytes},\ }\href@noop {} {\bibfield  {journal} {\bibinfo  {journal}
  {J. Fluid Mech.}\ }\textbf {\bibinfo {volume} {148}},\ \bibinfo {pages} {247}
  (\bibinfo {year} {1984})}\BibitemShut {NoStop}%
\bibitem [{\citenamefont {Anderson}(1989)}]{anderson1989colloid}%
  \BibitemOpen
  \bibfield  {author} {\bibinfo {author} {\bibfnamefont {J.~L.}\ \bibnamefont
  {Anderson}},\ }\bibfield  {title} {\bibinfo {title} {Colloid transport by
  interfacial forces},\ }\href@noop {} {\bibfield  {journal} {\bibinfo
  {journal} {Ann. Rev. Fluid Mech.}\ }\textbf {\bibinfo {volume} {21}},\
  \bibinfo {pages} {61} (\bibinfo {year} {1989})}\BibitemShut {NoStop}%
\bibitem [{\citenamefont {Shin}\ \emph {et~al.}(2016)\citenamefont {Shin},
  \citenamefont {Um}, \citenamefont {Sabass}, \citenamefont {Ault},
  \citenamefont {Rahimi}, \citenamefont {Warren},\ and\ \citenamefont
  {Stone}}]{shin2016size}%
  \BibitemOpen
  \bibfield  {author} {\bibinfo {author} {\bibfnamefont {S.}~\bibnamefont
  {Shin}}, \bibinfo {author} {\bibfnamefont {E.}~\bibnamefont {Um}}, \bibinfo
  {author} {\bibfnamefont {B.}~\bibnamefont {Sabass}}, \bibinfo {author}
  {\bibfnamefont {J.~T.}\ \bibnamefont {Ault}}, \bibinfo {author}
  {\bibfnamefont {M.}~\bibnamefont {Rahimi}}, \bibinfo {author} {\bibfnamefont
  {P.~B.}\ \bibnamefont {Warren}},\ and\ \bibinfo {author} {\bibfnamefont
  {H.~A.}\ \bibnamefont {Stone}},\ }\bibfield  {title} {\bibinfo {title}
  {Size-dependent control of colloid transport via solute gradients in dead-end
  channels},\ }\href@noop {} {\bibfield  {journal} {\bibinfo  {journal} {Proc.
  Nat. Acad. Sci.}\ }\textbf {\bibinfo {volume} {113}},\ \bibinfo {pages} {257}
  (\bibinfo {year} {2016})}\BibitemShut {NoStop}%
\bibitem [{\citenamefont {Velegol}\ \emph {et~al.}(2016)\citenamefont
  {Velegol}, \citenamefont {Garg}, \citenamefont {Guha}, \citenamefont {Kar},\
  and\ \citenamefont {Kumar}}]{velegol2016origins}%
  \BibitemOpen
  \bibfield  {author} {\bibinfo {author} {\bibfnamefont {D.}~\bibnamefont
  {Velegol}}, \bibinfo {author} {\bibfnamefont {A.}~\bibnamefont {Garg}},
  \bibinfo {author} {\bibfnamefont {R.}~\bibnamefont {Guha}}, \bibinfo {author}
  {\bibfnamefont {A.}~\bibnamefont {Kar}},\ and\ \bibinfo {author}
  {\bibfnamefont {M.}~\bibnamefont {Kumar}},\ }\bibfield  {title} {\bibinfo
  {title} {Origins of concentration gradients for diffusiophoresis},\
  }\href@noop {} {\bibfield  {journal} {\bibinfo  {journal} {Soft matter}\
  }\textbf {\bibinfo {volume} {12}},\ \bibinfo {pages} {4686} (\bibinfo {year}
  {2016})}\BibitemShut {NoStop}%
\bibitem [{\citenamefont {Banerjee}\ and\ \citenamefont
  {Squires}(2019)}]{banerjee2019long}%
  \BibitemOpen
  \bibfield  {author} {\bibinfo {author} {\bibfnamefont {A.}~\bibnamefont
  {Banerjee}}\ and\ \bibinfo {author} {\bibfnamefont {T.~M.}\ \bibnamefont
  {Squires}},\ }\bibfield  {title} {\bibinfo {title} {Long-range, selective,
  on-demand suspension interactions: Combining and triggering soluto-inertial
  beacons},\ }\href@noop {} {\bibfield  {journal} {\bibinfo  {journal} {Sci.
  Adv.}\ }\textbf {\bibinfo {volume} {5}},\ \bibinfo {pages} {eaax1893}
  (\bibinfo {year} {2019})}\BibitemShut {NoStop}%
\bibitem [{\citenamefont {Gupta}\ \emph {et~al.}(2019)\citenamefont {Gupta},
  \citenamefont {Rallabandi},\ and\ \citenamefont
  {Stone}}]{gupta2019diffusiophoretic}%
  \BibitemOpen
  \bibfield  {author} {\bibinfo {author} {\bibfnamefont {A.}~\bibnamefont
  {Gupta}}, \bibinfo {author} {\bibfnamefont {B.}~\bibnamefont {Rallabandi}},\
  and\ \bibinfo {author} {\bibfnamefont {H.~A.}\ \bibnamefont {Stone}},\
  }\bibfield  {title} {\bibinfo {title} {Diffusiophoretic and diffusioosmotic
  velocities for mixtures of valence-asymmetric electrolytes},\ }\href@noop {}
  {\bibfield  {journal} {\bibinfo  {journal} {Phys. Rev. Fluids}\ }\textbf
  {\bibinfo {volume} {4}},\ \bibinfo {pages} {043702} (\bibinfo {year}
  {2019})}\BibitemShut {NoStop}%
\bibitem [{\citenamefont {Gupta}\ \emph {et~al.}(2020)\citenamefont {Gupta},
  \citenamefont {Shim},\ and\ \citenamefont
  {Stone}}]{gupta2020diffusiophoresis}%
  \BibitemOpen
  \bibfield  {author} {\bibinfo {author} {\bibfnamefont {A.}~\bibnamefont
  {Gupta}}, \bibinfo {author} {\bibfnamefont {S.}~\bibnamefont {Shim}},\ and\
  \bibinfo {author} {\bibfnamefont {H.~A.}\ \bibnamefont {Stone}},\ }\bibfield
  {title} {\bibinfo {title} {Diffusiophoresis: from dilute to concentrated
  electrolytes},\ }\href@noop {} {\bibfield  {journal} {\bibinfo  {journal}
  {Soft Matter}\ }\textbf {\bibinfo {volume} {16}},\ \bibinfo {pages} {6975}
  (\bibinfo {year} {2020})}\BibitemShut {NoStop}%
\bibitem [{\citenamefont {Chu}\ \emph {et~al.}(2022)\citenamefont {Chu},
  \citenamefont {Garoff}, \citenamefont {Tilton},\ and\ \citenamefont
  {Khair}}]{chu2022tuning}%
  \BibitemOpen
  \bibfield  {author} {\bibinfo {author} {\bibfnamefont {H.~C.}\ \bibnamefont
  {Chu}}, \bibinfo {author} {\bibfnamefont {S.}~\bibnamefont {Garoff}},
  \bibinfo {author} {\bibfnamefont {R.~D.}\ \bibnamefont {Tilton}},\ and\
  \bibinfo {author} {\bibfnamefont {A.~S.}\ \bibnamefont {Khair}},\ }\bibfield
  {title} {\bibinfo {title} {Tuning chemotactic and diffusiophoretic spreading
  via hydrodynamic flows},\ }\href@noop {} {\bibfield  {journal} {\bibinfo
  {journal} {Soft Matter}\ }\textbf {\bibinfo {volume} {18}},\ \bibinfo {pages}
  {1896} (\bibinfo {year} {2022})}\BibitemShut {NoStop}%
\bibitem [{\citenamefont {Ganguly}\ and\ \citenamefont
  {Gupta}(2023)}]{ganguly2023going}%
  \BibitemOpen
  \bibfield  {author} {\bibinfo {author} {\bibfnamefont {A.}~\bibnamefont
  {Ganguly}}\ and\ \bibinfo {author} {\bibfnamefont {A.}~\bibnamefont
  {Gupta}},\ }\bibfield  {title} {\bibinfo {title} {Going in circles: Slender
  body analysis of a self-propelling bent rod},\ }\href@noop {} {\bibfield
  {journal} {\bibinfo  {journal} {Phys. Rev. Fluids}\ }\textbf {\bibinfo
  {volume} {8}},\ \bibinfo {pages} {014103} (\bibinfo {year}
  {2023})}\BibitemShut {NoStop}%
\bibitem [{\citenamefont {Raj}\ \emph {et~al.}(2023)\citenamefont {Raj},
  \citenamefont {Shields},\ and\ \citenamefont {Gupta}}]{raj2023two}%
  \BibitemOpen
  \bibfield  {author} {\bibinfo {author} {\bibfnamefont {R.~R.}\ \bibnamefont
  {Raj}}, \bibinfo {author} {\bibfnamefont {C.~W.}\ \bibnamefont {Shields}},\
  and\ \bibinfo {author} {\bibfnamefont {A.}~\bibnamefont {Gupta}},\ }\bibfield
   {title} {\bibinfo {title} {Two-dimensional diffusiophoretic colloidal
  banding: Optimizing the spatial and temporal design of solute sinks and
  sources},\ }\href@noop {} {\bibfield  {journal} {\bibinfo  {journal} {Soft
  Matter}\ }\textbf {\bibinfo {volume} {19}},\ \bibinfo {pages} {892} (\bibinfo
  {year} {2023})}\BibitemShut {NoStop}%
\bibitem [{\citenamefont {Shi}\ \emph {et~al.}(2016)\citenamefont {Shi},
  \citenamefont {Nery-Azevedo}, \citenamefont {Abdel-Fattah},\ and\
  \citenamefont {Squires}}]{shi2016diffusiophoretic}%
  \BibitemOpen
  \bibfield  {author} {\bibinfo {author} {\bibfnamefont {N.}~\bibnamefont
  {Shi}}, \bibinfo {author} {\bibfnamefont {R.}~\bibnamefont {Nery-Azevedo}},
  \bibinfo {author} {\bibfnamefont {A.~I.}\ \bibnamefont {Abdel-Fattah}},\ and\
  \bibinfo {author} {\bibfnamefont {T.~M.}\ \bibnamefont {Squires}},\
  }\bibfield  {title} {\bibinfo {title} {Diffusiophoretic focusing of suspended
  colloids},\ }\href@noop {} {\bibfield  {journal} {\bibinfo  {journal} {Phys.
  Rev. Lett.}\ }\textbf {\bibinfo {volume} {117}},\ \bibinfo {pages} {258001}
  (\bibinfo {year} {2016})}\BibitemShut {NoStop}%
\bibitem [{\citenamefont {Deen}(2012)}]{deen2012analysis}%
  \BibitemOpen
  \bibfield  {author} {\bibinfo {author} {\bibfnamefont {W.}~\bibnamefont
  {Deen}},\ }\href {https://books.google.com/books?id=60YsAwEACAAJ} {\emph
  {\bibinfo {title} {Analysis of Transport Phenomena}}},\ Topics in chemical
  engineering\ (\bibinfo  {publisher} {Oxford University Press},\ \bibinfo
  {year} {2012})\BibitemShut {NoStop}%
\bibitem [{\citenamefont {Bard}\ and\ \citenamefont
  {Faulkner}(2000)}]{bard2000electrochemical}%
  \BibitemOpen
  \bibfield  {author} {\bibinfo {author} {\bibfnamefont {A.}~\bibnamefont
  {Bard}}\ and\ \bibinfo {author} {\bibfnamefont {L.}~\bibnamefont
  {Faulkner}},\ }\href {https://books.google.com/books?id=hQocAAAAQBAJ} {\emph
  {\bibinfo {title} {Electrochemical Methods: Fundamentals and Applications,
  2nd Edition}}}\ (\bibinfo  {publisher} {John Wiley \& Sons, Incorporated},\
  \bibinfo {year} {2000})\BibitemShut {NoStop}%
\bibitem [{\citenamefont {Newman}\ and\ \citenamefont
  {Thomas-Alyea}(2012)}]{newman2012electrochemical}%
  \BibitemOpen
  \bibfield  {author} {\bibinfo {author} {\bibfnamefont {J.}~\bibnamefont
  {Newman}}\ and\ \bibinfo {author} {\bibfnamefont {K.}~\bibnamefont
  {Thomas-Alyea}},\ }\href {https://books.google.com/books?id=ewe4DQAAQBAJ}
  {\emph {\bibinfo {title} {Electrochemical Systems}}},\ The ECS Series of
  Texts and Monographs\ (\bibinfo  {publisher} {Wiley},\ \bibinfo {year}
  {2012})\BibitemShut {NoStop}%
\bibitem [{\citenamefont {Jarvey}\ \emph {et~al.}(2022)\citenamefont {Jarvey},
  \citenamefont {Henrique},\ and\ \citenamefont {Gupta}}]{jarvey2022ion}%
  \BibitemOpen
  \bibfield  {author} {\bibinfo {author} {\bibfnamefont {N.}~\bibnamefont
  {Jarvey}}, \bibinfo {author} {\bibfnamefont {F.}~\bibnamefont {Henrique}},\
  and\ \bibinfo {author} {\bibfnamefont {A.}~\bibnamefont {Gupta}},\ }\bibfield
   {title} {\bibinfo {title} {Ion transport in an electrochemical cell: A
  theoretical framework to couple dynamics of double layers and redox reactions
  for multicomponent electrolyte solutions},\ }\href@noop {} {\bibfield
  {journal} {\bibinfo  {journal} {J. Electrochem. Soc.}\ }\textbf {\bibinfo
  {volume} {169}},\ \bibinfo {pages} {093506} (\bibinfo {year}
  {2022})}\BibitemShut {NoStop}%
\bibitem [{\citenamefont {Henrique}\ \emph
  {et~al.}(2022{\natexlab{a}})\citenamefont {Henrique}, \citenamefont {Zuk},\
  and\ \citenamefont {Gupta}}]{henrique2022charging}%
  \BibitemOpen
  \bibfield  {author} {\bibinfo {author} {\bibfnamefont {F.}~\bibnamefont
  {Henrique}}, \bibinfo {author} {\bibfnamefont {P.~J.}\ \bibnamefont {Zuk}},\
  and\ \bibinfo {author} {\bibfnamefont {A.}~\bibnamefont {Gupta}},\ }\bibfield
   {title} {\bibinfo {title} {Charging dynamics of electrical double layers
  inside a cylindrical pore: predicting the effects of arbitrary pore size},\
  }\href@noop {} {\bibfield  {journal} {\bibinfo  {journal} {Soft Matter}\
  }\textbf {\bibinfo {volume} {18}},\ \bibinfo {pages} {198} (\bibinfo {year}
  {2022}{\natexlab{a}})}\BibitemShut {NoStop}%
\bibitem [{\citenamefont {Henrique}\ \emph
  {et~al.}(2022{\natexlab{b}})\citenamefont {Henrique}, \citenamefont {Zuk},\
  and\ \citenamefont {Gupta}}]{henrique2022impact}%
  \BibitemOpen
  \bibfield  {author} {\bibinfo {author} {\bibfnamefont {F.}~\bibnamefont
  {Henrique}}, \bibinfo {author} {\bibfnamefont {P.~J.}\ \bibnamefont {Zuk}},\
  and\ \bibinfo {author} {\bibfnamefont {A.}~\bibnamefont {Gupta}},\ }\bibfield
   {title} {\bibinfo {title} {Impact of asymmetries in valences and
  diffusivities on the transport of a binary electrolyte in a charged
  cylindrical pore},\ }\href@noop {} {\bibfield  {journal} {\bibinfo  {journal}
  {Electrochim. Acta}\ }\textbf {\bibinfo {volume} {433}},\ \bibinfo {pages}
  {141220} (\bibinfo {year} {2022}{\natexlab{b}})}\BibitemShut {NoStop}%
\bibitem [{\citenamefont {de~Levie}(1963)}]{de1963porous}%
  \BibitemOpen
  \bibfield  {author} {\bibinfo {author} {\bibfnamefont {R.}~\bibnamefont
  {de~Levie}},\ }\bibfield  {title} {\bibinfo {title} {On porous electrodes in
  electrolyte solutions: I. capacitance effects},\ }\href@noop {} {\bibfield
  {journal} {\bibinfo  {journal} {Electrochim. Acta}\ }\textbf {\bibinfo
  {volume} {8}},\ \bibinfo {pages} {751} (\bibinfo {year} {1963})}\BibitemShut
  {NoStop}%
\bibitem [{\citenamefont {Keh}\ and\ \citenamefont
  {Wei}(2000)}]{keh2000diffusiophoretic}%
  \BibitemOpen
  \bibfield  {author} {\bibinfo {author} {\bibfnamefont {H.~J.}\ \bibnamefont
  {Keh}}\ and\ \bibinfo {author} {\bibfnamefont {Y.~K.}\ \bibnamefont {Wei}},\
  }\bibfield  {title} {\bibinfo {title} {Diffusiophoretic mobility of spherical
  particles at low potential and arbitrary double-layer thickness},\
  }\href@noop {} {\bibfield  {journal} {\bibinfo  {journal} {Langmuir}\
  }\textbf {\bibinfo {volume} {16}},\ \bibinfo {pages} {5289} (\bibinfo {year}
  {2000})}\BibitemShut {NoStop}%
\bibitem [{\citenamefont {Hollingsworth}\ and\ \citenamefont
  {Saville}(2003)}]{hollingsworth2003broad}%
  \BibitemOpen
  \bibfield  {author} {\bibinfo {author} {\bibfnamefont {A.}~\bibnamefont
  {Hollingsworth}}\ and\ \bibinfo {author} {\bibfnamefont {D.}~\bibnamefont
  {Saville}},\ }\bibfield  {title} {\bibinfo {title} {A broad frequency range
  dielectric spectrometer for colloidal suspensions: cell design, calibration,
  and validation},\ }\href@noop {} {\bibfield  {journal} {\bibinfo  {journal}
  {J. Colloid Interface Sci.}\ }\textbf {\bibinfo {volume} {257}},\ \bibinfo
  {pages} {65} (\bibinfo {year} {2003})}\BibitemShut {NoStop}%
\bibitem [{\citenamefont {Krozel}\ and\ \citenamefont
  {Saville}(1992)}]{krozel1992electrostatic}%
  \BibitemOpen
  \bibfield  {author} {\bibinfo {author} {\bibfnamefont {J.}~\bibnamefont
  {Krozel}}\ and\ \bibinfo {author} {\bibfnamefont {D.}~\bibnamefont
  {Saville}},\ }\bibfield  {title} {\bibinfo {title} {Electrostatic
  interactions between two spheres: Solutions of the debye-h{\"u}ckel equation
  with a charge regulation boundary condition},\ }\href@noop {} {\bibfield
  {journal} {\bibinfo  {journal} {J. Colloid Interface Sci.}\ }\textbf
  {\bibinfo {volume} {150}},\ \bibinfo {pages} {365} (\bibinfo {year}
  {1992})}\BibitemShut {NoStop}%
\bibitem [{\citenamefont {Balu}\ and\ \citenamefont
  {Khair}(2022)}]{balu2022electrochemical}%
  \BibitemOpen
  \bibfield  {author} {\bibinfo {author} {\bibfnamefont {B.}~\bibnamefont
  {Balu}}\ and\ \bibinfo {author} {\bibfnamefont {A.~S.}\ \bibnamefont
  {Khair}},\ }\bibfield  {title} {\bibinfo {title} {The electrochemical
  impedance spectrum of asymmetric electrolytes across low to moderate
  frequencies},\ }\href@noop {} {\bibfield  {journal} {\bibinfo  {journal} {J.
  Electroanal. Chem.}\ }\textbf {\bibinfo {volume} {911}},\ \bibinfo {pages}
  {116222} (\bibinfo {year} {2022})}\BibitemShut {NoStop}%
\bibitem [{\citenamefont {Haynes}(2016)}]{haynes2016crc}%
  \BibitemOpen
  \bibfield  {author} {\bibinfo {author} {\bibfnamefont {W.~M.}\ \bibnamefont
  {Haynes}},\ }\href@noop {} {\emph {\bibinfo {title} {CRC Handbook of
  Chemistry and Physics}}}\ (\bibinfo  {publisher} {CRC press},\ \bibinfo
  {year} {2016})\BibitemShut {NoStop}%
\bibitem [{\citenamefont {Bazant}\ \emph {et~al.}(2004)\citenamefont {Bazant},
  \citenamefont {Thornton},\ and\ \citenamefont {Ajdari}}]{bazant2004diffuse}%
  \BibitemOpen
  \bibfield  {author} {\bibinfo {author} {\bibfnamefont {M.~Z.}\ \bibnamefont
  {Bazant}}, \bibinfo {author} {\bibfnamefont {K.}~\bibnamefont {Thornton}},\
  and\ \bibinfo {author} {\bibfnamefont {A.}~\bibnamefont {Ajdari}},\
  }\bibfield  {title} {\bibinfo {title} {Diffuse-charge dynamics in
  electrochemical systems},\ }\href@noop {} {\bibfield  {journal} {\bibinfo
  {journal} {Phys. Rev. E}\ }\textbf {\bibinfo {volume} {70}},\ \bibinfo
  {pages} {021506} (\bibinfo {year} {2004})}\BibitemShut {NoStop}%
\bibitem [{\citenamefont {Biesheuvel}\ \emph {et~al.}(2011)\citenamefont
  {Biesheuvel}, \citenamefont {Fu},\ and\ \citenamefont
  {Bazant}}]{biesheuvel2011diffuse}%
  \BibitemOpen
  \bibfield  {author} {\bibinfo {author} {\bibfnamefont {P.}~\bibnamefont
  {Biesheuvel}}, \bibinfo {author} {\bibfnamefont {Y.}~\bibnamefont {Fu}},\
  and\ \bibinfo {author} {\bibfnamefont {M.~Z.}\ \bibnamefont {Bazant}},\
  }\bibfield  {title} {\bibinfo {title} {Diffuse charge and faradaic reactions
  in porous electrodes},\ }\href@noop {} {\bibfield  {journal} {\bibinfo
  {journal} {Phys. Rev. E}\ }\textbf {\bibinfo {volume} {83}},\ \bibinfo
  {pages} {061507} (\bibinfo {year} {2011})}\BibitemShut {NoStop}%
\bibitem [{\citenamefont {Balu}\ and\ \citenamefont
  {S. Khair}(2018)}]{balu_role_2018}%
  \BibitemOpen
  \bibfield  {author} {\bibinfo {author} {\bibfnamefont {B.}~\bibnamefont
  {Balu}}\ and\ \bibinfo {author} {\bibfnamefont {A.}~\bibnamefont
  {S. Khair}},\ }\bibfield  {title} {\bibinfo {title} {Role of
  {Stefan}–{Maxwell} fluxes in the dynamics of concentrated electrolytes},\
  }\href {https://doi.org/10.1039/C8SM01222A} {\bibfield  {journal} {\bibinfo
  {journal} {Soft Matter}\ }\textbf {\bibinfo {volume} {14}},\ \bibinfo {pages}
  {8267} (\bibinfo {year} {2018})},\ \bibinfo {note} {publisher: Royal Society
  of Chemistry}\BibitemShut {NoStop}%
\bibitem [{\citenamefont {Bazant}\ \emph {et~al.}(2005)\citenamefont {Bazant},
  \citenamefont {Chu},\ and\ \citenamefont {Bayly}}]{bazant2005current}%
  \BibitemOpen
  \bibfield  {author} {\bibinfo {author} {\bibfnamefont {M.~Z.}\ \bibnamefont
  {Bazant}}, \bibinfo {author} {\bibfnamefont {K.~T.}\ \bibnamefont {Chu}},\
  and\ \bibinfo {author} {\bibfnamefont {B.~J.}\ \bibnamefont {Bayly}},\
  }\bibfield  {title} {\bibinfo {title} {Current-voltage relations for
  electrochemical thin films},\ }\href@noop {} {\bibfield  {journal} {\bibinfo
  {journal} {SIAM J. Appl. Math.}\ }\textbf {\bibinfo {volume} {65}},\ \bibinfo
  {pages} {1463} (\bibinfo {year} {2005})}\BibitemShut {NoStop}%
\bibitem [{\citenamefont {Mittal}\ and\ \citenamefont
  {Furst}(2009)}]{mittal2009electric}%
  \BibitemOpen
  \bibfield  {author} {\bibinfo {author} {\bibfnamefont {M.}~\bibnamefont
  {Mittal}}\ and\ \bibinfo {author} {\bibfnamefont {E.~M.}\ \bibnamefont
  {Furst}},\ }\bibfield  {title} {\bibinfo {title} {Electric field-directed
  convective assembly of ellipsoidal colloidal particles to create optically
  and mechanically anisotropic thin films},\ }\href@noop {} {\bibfield
  {journal} {\bibinfo  {journal} {Adv. Funtc. Mat.}\ }\textbf {\bibinfo
  {volume} {19}},\ \bibinfo {pages} {3271} (\bibinfo {year}
  {2009})}\BibitemShut {NoStop}%
\bibitem [{\citenamefont {Demir{\"o}rs}\ \emph {et~al.}(2010)\citenamefont
  {Demir{\"o}rs}, \citenamefont {Johnson}, \citenamefont {van Kats},
  \citenamefont {van Blaaderen},\ and\ \citenamefont
  {Imhof}}]{demirors2010directed}%
  \BibitemOpen
  \bibfield  {author} {\bibinfo {author} {\bibfnamefont {A.~F.}\ \bibnamefont
  {Demir{\"o}rs}}, \bibinfo {author} {\bibfnamefont {P.~M.}\ \bibnamefont
  {Johnson}}, \bibinfo {author} {\bibfnamefont {C.~M.}\ \bibnamefont {van
  Kats}}, \bibinfo {author} {\bibfnamefont {A.}~\bibnamefont {van Blaaderen}},\
  and\ \bibinfo {author} {\bibfnamefont {A.}~\bibnamefont {Imhof}},\ }\bibfield
   {title} {\bibinfo {title} {Directed self-assembly of colloidal dumbbells
  with an electric field},\ }\href@noop {} {\bibfield  {journal} {\bibinfo
  {journal} {Langmuir}\ }\textbf {\bibinfo {volume} {26}},\ \bibinfo {pages}
  {14466} (\bibinfo {year} {2010})}\BibitemShut {NoStop}%
\bibitem [{\citenamefont {Ma}\ \emph {et~al.}(2012)\citenamefont {Ma},
  \citenamefont {Wang}, \citenamefont {Smith},\ and\ \citenamefont
  {Wu}}]{ma2012two}%
  \BibitemOpen
  \bibfield  {author} {\bibinfo {author} {\bibfnamefont {F.}~\bibnamefont
  {Ma}}, \bibinfo {author} {\bibfnamefont {S.}~\bibnamefont {Wang}}, \bibinfo
  {author} {\bibfnamefont {L.}~\bibnamefont {Smith}},\ and\ \bibinfo {author}
  {\bibfnamefont {N.}~\bibnamefont {Wu}},\ }\bibfield  {title} {\bibinfo
  {title} {Two-dimensional assembly of symmetric colloidal dimers under
  electric fields},\ }\href@noop {} {\bibfield  {journal} {\bibinfo  {journal}
  {Adv. Funct. Mat.}\ }\textbf {\bibinfo {volume} {22}},\ \bibinfo {pages}
  {4334} (\bibinfo {year} {2012})}\BibitemShut {NoStop}%
\bibitem [{\citenamefont {Panczyk}\ \emph {et~al.}(2013)\citenamefont
  {Panczyk}, \citenamefont {Park}, \citenamefont {Wagner},\ and\ \citenamefont
  {Furst}}]{panczyk2013two}%
  \BibitemOpen
  \bibfield  {author} {\bibinfo {author} {\bibfnamefont {M.~M.}\ \bibnamefont
  {Panczyk}}, \bibinfo {author} {\bibfnamefont {J.-G.}\ \bibnamefont {Park}},
  \bibinfo {author} {\bibfnamefont {N.~J.}\ \bibnamefont {Wagner}},\ and\
  \bibinfo {author} {\bibfnamefont {E.~M.}\ \bibnamefont {Furst}},\ }\bibfield
  {title} {\bibinfo {title} {Two-dimensional directed assembly of dicolloids},\
  }\href@noop {} {\bibfield  {journal} {\bibinfo  {journal} {Langmuir}\
  }\textbf {\bibinfo {volume} {29}},\ \bibinfo {pages} {75} (\bibinfo {year}
  {2013})}\BibitemShut {NoStop}%
\bibitem [{\citenamefont {Zhang}\ and\ \citenamefont
  {Zhu}(2012)}]{zhang2012directed}%
  \BibitemOpen
  \bibfield  {author} {\bibinfo {author} {\bibfnamefont {L.}~\bibnamefont
  {Zhang}}\ and\ \bibinfo {author} {\bibfnamefont {Y.}~\bibnamefont {Zhu}},\
  }\bibfield  {title} {\bibinfo {title} {Directed assembly of janus particles
  under high frequency ac-electric fields: Effects of medium conductivity and
  colloidal surface chemistry},\ }\href@noop {} {\bibfield  {journal} {\bibinfo
   {journal} {Langmuir}\ }\textbf {\bibinfo {volume} {28}},\ \bibinfo {pages}
  {13201} (\bibinfo {year} {2012})}\BibitemShut {NoStop}%
\bibitem [{\citenamefont {Shields~IV}\ \emph {et~al.}(2013)\citenamefont
  {Shields~IV}, \citenamefont {Zhu}, \citenamefont {Yang}, \citenamefont
  {Bharti}, \citenamefont {Liu}, \citenamefont {Yellen}, \citenamefont
  {Velev},\ and\ \citenamefont {L{\'o}pez}}]{shields2013field}%
  \BibitemOpen
  \bibfield  {author} {\bibinfo {author} {\bibfnamefont {C.~W.}\ \bibnamefont
  {Shields~IV}}, \bibinfo {author} {\bibfnamefont {S.}~\bibnamefont {Zhu}},
  \bibinfo {author} {\bibfnamefont {Y.}~\bibnamefont {Yang}}, \bibinfo {author}
  {\bibfnamefont {B.}~\bibnamefont {Bharti}}, \bibinfo {author} {\bibfnamefont
  {J.}~\bibnamefont {Liu}}, \bibinfo {author} {\bibfnamefont {B.~B.}\
  \bibnamefont {Yellen}}, \bibinfo {author} {\bibfnamefont {O.~D.}\
  \bibnamefont {Velev}},\ and\ \bibinfo {author} {\bibfnamefont {G.~P.}\
  \bibnamefont {L{\'o}pez}},\ }\bibfield  {title} {\bibinfo {title}
  {Field-directed assembly of patchy anisotropic microparticles with defined
  shape},\ }\href@noop {} {\bibfield  {journal} {\bibinfo  {journal} {Soft
  Matter}\ }\textbf {\bibinfo {volume} {9}},\ \bibinfo {pages} {9219} (\bibinfo
  {year} {2013})}\BibitemShut {NoStop}%
\bibitem [{\citenamefont {Ma}\ \emph {et~al.}(2015)\citenamefont {Ma},
  \citenamefont {Wang}, \citenamefont {Wu},\ and\ \citenamefont
  {Wu}}]{ma2015electric}%
  \BibitemOpen
  \bibfield  {author} {\bibinfo {author} {\bibfnamefont {F.}~\bibnamefont
  {Ma}}, \bibinfo {author} {\bibfnamefont {S.}~\bibnamefont {Wang}}, \bibinfo
  {author} {\bibfnamefont {D.~T.}\ \bibnamefont {Wu}},\ and\ \bibinfo {author}
  {\bibfnamefont {N.}~\bibnamefont {Wu}},\ }\bibfield  {title} {\bibinfo
  {title} {Electric-field--induced assembly and propulsion of chiral colloidal
  clusters},\ }\href@noop {} {\bibfield  {journal} {\bibinfo  {journal} {Proc.
  Natl. Acad. Sci.}\ }\textbf {\bibinfo {volume} {112}},\ \bibinfo {pages}
  {6307} (\bibinfo {year} {2015})}\BibitemShut {NoStop}%
\bibitem [{\citenamefont {Kruglova}\ \emph {et~al.}(2013)\citenamefont
  {Kruglova}, \citenamefont {Demeyer}, \citenamefont {Zhong}, \citenamefont
  {Zhou},\ and\ \citenamefont {Clays}}]{kruglova_wonders_2013}%
  \BibitemOpen
  \bibfield  {author} {\bibinfo {author} {\bibfnamefont {O.}~\bibnamefont
  {Kruglova}}, \bibinfo {author} {\bibfnamefont {P.-J.}\ \bibnamefont
  {Demeyer}}, \bibinfo {author} {\bibfnamefont {K.}~\bibnamefont {Zhong}},
  \bibinfo {author} {\bibfnamefont {Y.}~\bibnamefont {Zhou}},\ and\ \bibinfo
  {author} {\bibfnamefont {K.}~\bibnamefont {Clays}},\ }\bibfield  {title}
  {\bibinfo {title} {Wonders of colloidal assembly},\ }\href
  {https://doi.org/10.1039/C3SM50845E} {\bibfield  {journal} {\bibinfo
  {journal} {Soft Matter}\ }\textbf {\bibinfo {volume} {9}},\ \bibinfo {pages}
  {9072} (\bibinfo {year} {2013})},\ \bibinfo {note} {publisher: Royal Society
  of Chemistry}\BibitemShut {NoStop}%
\bibitem [{\citenamefont {Li}\ \emph {et~al.}(2018)\citenamefont {Li},
  \citenamefont {Lu}, \citenamefont {Chen},\ and\ \citenamefont
  {Amine}}]{li201830}%
  \BibitemOpen
  \bibfield  {author} {\bibinfo {author} {\bibfnamefont {M.}~\bibnamefont
  {Li}}, \bibinfo {author} {\bibfnamefont {J.}~\bibnamefont {Lu}}, \bibinfo
  {author} {\bibfnamefont {Z.}~\bibnamefont {Chen}},\ and\ \bibinfo {author}
  {\bibfnamefont {K.}~\bibnamefont {Amine}},\ }\bibfield  {title} {\bibinfo
  {title} {30 years of lithium-ion batteries},\ }\href@noop {} {\bibfield
  {journal} {\bibinfo  {journal} {Adv. Mater.}\ }\textbf {\bibinfo {volume}
  {30}},\ \bibinfo {pages} {1800561} (\bibinfo {year} {2018})}\BibitemShut
  {NoStop}%
\bibitem [{\citenamefont {Zhang}(2006)}]{zhang2006review}%
  \BibitemOpen
  \bibfield  {author} {\bibinfo {author} {\bibfnamefont {S.~S.}\ \bibnamefont
  {Zhang}},\ }\bibfield  {title} {\bibinfo {title} {A review on electrolyte
  additives for lithium-ion batteries},\ }\href@noop {} {\bibfield  {journal}
  {\bibinfo  {journal} {J. Power Sources}\ }\textbf {\bibinfo {volume} {162}},\
  \bibinfo {pages} {1379} (\bibinfo {year} {2006})}\BibitemShut {NoStop}%
\bibitem [{\citenamefont {Biesheuvel}\ \emph {et~al.}(2012)\citenamefont
  {Biesheuvel}, \citenamefont {Fu},\ and\ \citenamefont
  {Bazant}}]{biesheuvel_electrochemistry_2012}%
  \BibitemOpen
  \bibfield  {author} {\bibinfo {author} {\bibfnamefont {P.~M.}\ \bibnamefont
  {Biesheuvel}}, \bibinfo {author} {\bibfnamefont {Y.}~\bibnamefont {Fu}},\
  and\ \bibinfo {author} {\bibfnamefont {M.~Z.}\ \bibnamefont {Bazant}},\
  }\bibfield  {title} {\bibinfo {title} {Electrochemistry and capacitive
  charging of porous electrodes in asymmetric multicomponent electrolytes},\
  }\href {https://doi.org/10.1134/S1023193512060031} {\bibfield  {journal}
  {\bibinfo  {journal} {Russ. J. Electrochem.}\ }\textbf {\bibinfo {volume}
  {48}},\ \bibinfo {pages} {580} (\bibinfo {year} {2012})}\BibitemShut
  {NoStop}%
\bibitem [{\citenamefont {Sayed}\ \emph {et~al.}(2021)\citenamefont {Sayed},
  \citenamefont {Al~Radi}, \citenamefont {Ahmad}, \citenamefont {Abdelkareem},
  \citenamefont {Alawadhi}, \citenamefont {Atieh},\ and\ \citenamefont
  {Olabi}}]{sayed_faradic_2021}%
  \BibitemOpen
  \bibfield  {author} {\bibinfo {author} {\bibfnamefont {E.~T.}\ \bibnamefont
  {Sayed}}, \bibinfo {author} {\bibfnamefont {M.}~\bibnamefont {Al~Radi}},
  \bibinfo {author} {\bibfnamefont {A.}~\bibnamefont {Ahmad}}, \bibinfo
  {author} {\bibfnamefont {M.~A.}\ \bibnamefont {Abdelkareem}}, \bibinfo
  {author} {\bibfnamefont {H.}~\bibnamefont {Alawadhi}}, \bibinfo {author}
  {\bibfnamefont {M.~A.}\ \bibnamefont {Atieh}},\ and\ \bibinfo {author}
  {\bibfnamefont {A.~G.}\ \bibnamefont {Olabi}},\ }\bibfield  {title} {\bibinfo
  {title} {Faradic capacitive deionization ({FCDI}) for desalination and ion
  removal from wastewater},\ }\href
  {https://doi.org/10.1016/j.chemosphere.2021.130001} {\bibfield  {journal}
  {\bibinfo  {journal} {Chemosphere}\ }\textbf {\bibinfo {volume} {275}},\
  \bibinfo {pages} {130001} (\bibinfo {year} {2021})}\BibitemShut {NoStop}%
\bibitem [{\citenamefont {Wu}\ \emph {et~al.}(2017)\citenamefont {Wu},
  \citenamefont {Huang}, \citenamefont {Ye},\ and\ \citenamefont
  {Li}}]{wu2017co2}%
  \BibitemOpen
  \bibfield  {author} {\bibinfo {author} {\bibfnamefont {J.}~\bibnamefont
  {Wu}}, \bibinfo {author} {\bibfnamefont {Y.}~\bibnamefont {Huang}}, \bibinfo
  {author} {\bibfnamefont {W.}~\bibnamefont {Ye}},\ and\ \bibinfo {author}
  {\bibfnamefont {Y.}~\bibnamefont {Li}},\ }\bibfield  {title} {\bibinfo
  {title} {Co2 reduction: from the electrochemical to photochemical approach},\
  }\href@noop {} {\bibfield  {journal} {\bibinfo  {journal} {Adv. Sci,}\
  }\textbf {\bibinfo {volume} {4}},\ \bibinfo {pages} {1700194} (\bibinfo
  {year} {2017})}\BibitemShut {NoStop}%
\bibitem [{\citenamefont {Simon}\ and\ \citenamefont
  {Gogotsi}(2020)}]{simon2020perspectives}%
  \BibitemOpen
  \bibfield  {author} {\bibinfo {author} {\bibfnamefont {P.}~\bibnamefont
  {Simon}}\ and\ \bibinfo {author} {\bibfnamefont {Y.}~\bibnamefont
  {Gogotsi}},\ }\bibfield  {title} {\bibinfo {title} {Perspectives for
  electrochemical capacitors and related devices},\ }\href@noop {} {\bibfield
  {journal} {\bibinfo  {journal} {Nat. Mater.}\ }\textbf {\bibinfo {volume}
  {19}},\ \bibinfo {pages} {1151} (\bibinfo {year} {2020})}\BibitemShut
  {NoStop}%
\bibitem [{\citenamefont {Tao}\ \emph {et~al.}(2021)\citenamefont {Tao},
  \citenamefont {Liu}, \citenamefont {Yu},\ and\ \citenamefont
  {Cheng}}]{tao2021reversible}%
  \BibitemOpen
  \bibfield  {author} {\bibinfo {author} {\bibfnamefont {X.}~\bibnamefont
  {Tao}}, \bibinfo {author} {\bibfnamefont {D.}~\bibnamefont {Liu}}, \bibinfo
  {author} {\bibfnamefont {J.}~\bibnamefont {Yu}},\ and\ \bibinfo {author}
  {\bibfnamefont {H.}~\bibnamefont {Cheng}},\ }\bibfield  {title} {\bibinfo
  {title} {Reversible metal electrodeposition devices: an emerging approach to
  effective light modulation and thermal management},\ }\href@noop {}
  {\bibfield  {journal} {\bibinfo  {journal} {Adv. Opt. Mater.}\ }\textbf
  {\bibinfo {volume} {9}},\ \bibinfo {pages} {2001847} (\bibinfo {year}
  {2021})}\BibitemShut {NoStop}%
\bibitem [{\citenamefont {de~Levie}(1964)}]{de1964porous}%
  \BibitemOpen
  \bibfield  {author} {\bibinfo {author} {\bibfnamefont {R.}~\bibnamefont
  {de~Levie}},\ }\bibfield  {title} {\bibinfo {title} {On porous electrodes in
  electrolyte solutions—iv},\ }\href@noop {} {\bibfield  {journal} {\bibinfo
  {journal} {Electrochim. Acta}\ }\textbf {\bibinfo {volume} {9}},\ \bibinfo
  {pages} {1231} (\bibinfo {year} {1964})}\BibitemShut {NoStop}%
\bibitem [{\citenamefont {He}\ \emph {et~al.}(2018)\citenamefont {He},
  \citenamefont {Biesheuvel}, \citenamefont {Bazant},\ and\ \citenamefont
  {Hatton}}]{he2018theory}%
  \BibitemOpen
  \bibfield  {author} {\bibinfo {author} {\bibfnamefont {F.}~\bibnamefont
  {He}}, \bibinfo {author} {\bibfnamefont {P.}~\bibnamefont {Biesheuvel}},
  \bibinfo {author} {\bibfnamefont {M.~Z.}\ \bibnamefont {Bazant}},\ and\
  \bibinfo {author} {\bibfnamefont {T.~A.}\ \bibnamefont {Hatton}},\ }\bibfield
   {title} {\bibinfo {title} {Theory of water treatment by capacitive
  deionization with redox active porous electrodes},\ }\href@noop {} {\bibfield
   {journal} {\bibinfo  {journal} {Water Res.}\ }\textbf {\bibinfo {volume}
  {132}},\ \bibinfo {pages} {282} (\bibinfo {year} {2018})}\BibitemShut
  {NoStop}%
\end{thebibliography}%

\end{document}